\documentclass{iucrjournals}

\usepackage{amsmath,amssymb,mathtools}
\usepackage{siunitx}
\DeclareSIUnit\angstrom{\text{\AA}}
\usepackage{longtable}
\usepackage{supertabular}
\usepackage{microtype}
\usepackage{flafter}

\title{Updated all-electron Dirac--Fock densities and an element-adaptive parameterisation of scattering factors and potentials for neutral atoms}

\author[a,c,e]{I.~Lobato\IUCrCemaillink{ivan.lobato@neuralsoftx.com}}%
\author[b]{Z.~Zhang}%
\author[c,d]{S.~Van Aert}%
\author[e,f,g]{A.~I. Kirkland\IUCrCemaillink{angus.kirkland@materials.ox.ac.uk}}

\affil[a]{NeuralSoftX, Antverpiastraat 31, 2660 Antwerp, Belgium}
\affil[b]{AI for Science Institute, 150 Chengfu Road, Haidian District, Beijing 100080, China}
\affil[c]{Electron Microscopy for Materials Science (EMAT), University of Antwerp, Groenenborgerlaan 171, 2020 Antwerp, Belgium}
\affil[d]{NANOlight Centre of Excellence, University of Antwerp, Groenenborgerlaan 171, 2020 Antwerp, Belgium}
\affil[e]{Department of Materials, University of Oxford, Parks Road, Oxford OX1 3PH, United Kingdom}
\affil[f]{Rosalind Franklin Institute, Building R113, Rutherford Appleton Laboratory, Harwell Campus, Didcot, Oxfordshire OX11 0QX, United Kingdom}
\affil[g]{Electron Physical Science Imaging Centre, Diamond Light Source Ltd., Diamond House, OX11 0DE, United Kingdom}

\begin{document}
\nolinenumbers
\maketitle
\raggedbottom

\begin{synopsis}
Updated relativistic reference densities and an element-adaptive analytic parameterisation provide consistent electron and X-ray scattering factors, electron densities and electrostatic potentials for all 118 neutral atoms. The new basis improves reciprocal-space accuracy by three to four orders of magnitude over a controlled five-term refit while retaining closed-form derived quantities.
\end{synopsis}

\begin{abstract}
Updated reference data and an analytic parameterisation of elastic electron and X-ray scattering are presented for all 118 neutral atoms. The reference electron densities for the multi-electron elements $Z=2$--$118$ are computed with the relativistic B-spline Dirac--Fock code \texttt{atomx}, while hydrogen is constructed from the exact relativistic one-electron Dirac $1s$ solution; the electron scattering factor $f_e(g)$, X-ray scattering factor $f_x(g)$ and radial moments are derived from these densities. The reported parameterisation extends the fixed-size Lobato--Van Dyck hydrogenic expansion while retaining closed-form expressions for $f_x(g)$, $\rho(r)$, the electrostatic potential $V(r)$ and the projected potential $V(R)$. These extensions are an element-adaptive basis size $n_t(Z)$, a simultaneous real- and reciprocal-space fit, an exact $\langle r^4\rangle$ constraint in place of the non-relativistic Kato cusp, and a charge-carrying Dirac--Pad\'e basis term that adds a polynomial-times-exponential shape channel without replacing the hydrogenic basis by a tabulated Dirac radial function or assigning the term to a physical shell. For the same total parameter count, the Dirac--Pad\'e-enriched basis improves on the parameter-matched non-relativistic basis for 111 of the 118 elements, lowering the mean total cost by \SI{39}{\percent}. Relative to a controlled fixed five-term refit on the same reference grid and objective, the element-adaptive bases improve the median reciprocal-space deviations by about three to four orders of magnitude and resolve shell structure in $4\pi r^2\rho(r)$ that the fixed five-term basis cannot. The largest changes occur near the nucleus and in the reciprocal-space tail beyond the legacy \SI{12}{\per\angstrom} range, which is directly relevant to quantitative high-angle scattering and electron-diffraction measurements.
\end{abstract}


\section{Introduction}
\label{sec:intro}

Quantitative electron microscopy relies on accurate simulations of high-energy electron scattering from the electrostatic potential of a specimen. For high-resolution imaging, convergent-beam electron diffraction (CBED), four-dimensional scanning transmission electron microscopy (4D-STEM) and ptychography, the forward model that links structure to a measured signal is obtained by solving the high-energy Schr\"odinger equation. In practice this is achieved through the multislice method \cite{Cowley1957,VanDyck1985,Lobato2012} or the Bloch-wave formalism \cite{Bethe1928}. In almost every case the total electrostatic potential is assembled from atomic contributions using the independent atom model (IAM). In the IAM, the specimen electrostatic potential is a linear superposition of spherically symmetric atomic electrostatic potentials, each determined by the point nucleus and the surrounding electron density through Poisson's equation. The Fourier transform of the electron density is the X-ray scattering factor $f_x(g)$, and the Mott--Bethe relation \cite{Bethe1930,Mott1930} converts this into an electron scattering factor $f_e(g)$, with the point nuclear charge providing the leading high-$g$ term. The electron density that underpins all of these quantities is obtained from atomic-structure theory through self-consistent solutions of the relativistic Dirac equation, ranging from the early Dirac--Slater tabulations \cite{CromerWaber1965} to modern multi-configuration Dirac--Fock calculations \cite{Guerra2017}.

Tabulated numerical values of the X-ray and electron scattering factors are available across the periodic table \cite{CromerWaber1965,Kirkland1998}, with relativistic Hartree--Fock values and parametric fits for a representative set of atoms and ions \cite{DoyleTurner1968}. Simulation programs can use these values directly by interpolation, but analytic basis functions reduce storage requirements and, more importantly, enable the analytic computation of derived real-space quantities such as the electron density $\rho(r)$, the electrostatic potential $V(r)$, and the projected potential $V(R)$. All of these are needed for electron-scattering simulation and analysis using multislice and Bloch-wave calculations, partial-wave (Mott) cross-section codes, and electron-density analysis.

Several alternative parameterisations have been proposed \cite{SmithBurge1962,DoyleTurner1968,Peng1996} using sums of Gaussian functions. Such sums can reproduce the finite reciprocal-space interval over which they are fitted, but every Gaussian term decays exponentially in $g^2$ and a finite sum therefore cannot reproduce the $1/g^2$ asymptote required by the Mott--Bethe formula as $g\to\infty$. \citeasnoun{WeickenmeierKohl1991} altered each term as $[1-\exp(-B_i g^2)]/g^2$, which gives the electron scattering factor the correct $1/g^2$ asymptotic decay; through the Mott--Bethe relation, however, this makes the X-ray scattering factor a sum of Gaussians, so $f_x(g)$ decays too fast to reproduce the power-law tail of the true X-ray scattering factor. \citeasnoun{Kirkland1998} combined three Lorentzians and three Gaussians, $f_e(g)=\sum_i a_i/(g^2+b_i)+\sum_i c_i\exp(-d_i g^2)$, the Lorentzian terms $a_i/(g^2+b_i)$ giving $f_e$ the correct $1/g^2$ tail while the Gaussians add flexibility over the fitted interval. A systematic comparison of these parameterisations \cite{Lobato2015} showed that differences in the parameterised scattering factors propagate to measurable differences in simulated diffraction patterns, particularly at large $g$.

Fundamentally, all of these are fits to discrete, band-limited tabulations of $f_e(g)$ for which the high-$g$ behaviour is extrapolated rather than constrained. Hence, the X-ray scattering factor recovered through the inverse Mott--Bethe relation, $f_x(g) = Z - 2\pi^2 a_0\,g^2 f_e(g)$ (the inversion of equation~\eqref{eq:feg} below, with $a_0$ the Bohr radius), is unreliable because the explicit $g^2$ factor amplifies errors in the high-$g$ tail. The purely Gaussian forms then give $f_x(g)\to Z$ as $g\to\infty$ instead of the required $f_x(g)\to 0$, while the mixed Lorentzian--Gaussian form leaves $f_x$---and hence $\rho(r)$ and $V(r)$---with $g^2$-amplified artefacts because its amplitudes are not tied to this X-ray asymptotic constraint \cite{Lobato2014}. A consistent set of derived quantities therefore requires a basis whose asymptotics are correct by construction and whose coefficients satisfy the exact constraints.

The parameterisation of \citeasnoun{Lobato2014} used the analytic non-relativistic hydrogen electron scattering factor as the basis function. This choice has two key advantages. First, the hydrogen scattering factor naturally includes the correct $1/g^2$ asymptotic decay at large $g$, since it is derived from a physical Coulomb potential. Second, it yields closed-form analytic expressions for the X-ray scattering factor $f_x(g)$, the electron density $\rho(r)$, the electrostatic potential $V(r)$, and the projected potential $V(R)$. That parameterisation was shown to be one order of magnitude more accurate than the alternative analytic fits described above.

However, that parameterisation has three limitations that are addressed in the present work.

\textit{Fixed basis size.} The number of basis functions was fixed at $n_t = 5$ for all 103 elements. For H and He, five basis functions introduce unnecessary flexibility, whereas a fixed five-term basis becomes progressively insufficient as additional principal shells are occupied. This establishes the need for an element-adaptive basis size rather than a single global value.

\textit{Incomplete fitting target.} The original cost function fitted only the reciprocal-space scattering factors $f_e(g)$ and $f_x(g)$. These were subject to three exact analytic constraints (the charge-neutrality condition $f_x(0)=Z$, the Ibers relation for $f_e(0)$, and the Kato cusp condition \cite{Kato1957,Steiner1963}) and positivity of all derived quantities. However, the electron density $\rho(r)$ itself was not a fitting target. Density errors and errors in the near-nuclear electrostatic potential were therefore controlled only indirectly by sampled reciprocal-space data. This limitation was acknowledged in the original paper, where we noted that ``it would be better to parameterize the numerical radial atomic potential \ldots\ this requires knowledge of the tabulated values of radial electron-density distribution for all the atoms'' \cite{Lobato2014}. Moreover, the reciprocal-space target was sampled only to about \SI{12}{\per\angstrom}, the conventional range of tabulated scattering factors. The high-$g$ tail, which governs the near-nuclear electrostatic potential, was therefore constrained by the analytic form of the basis rather than directly by reference data. The present work makes the electron density an explicit fitting target, so that electron shell structure is constrained directly rather than only through a sampled transform.

\textit{Kato cusp condition.} The Kato cusp condition \cite{Kato1957,Steiner1963} was imposed as an exact constraint on the parameterised density. That condition was derived for non-relativistic point-nucleus calculations; the relativistic electron density does not exhibit a cusp at the nuclear position when a finite-sized nuclear charge distribution is used \cite{Mastalerz2010}. Direct access to a reference electron density makes this condition both unnecessary and inappropriate.

In this work we first compute updated all-electron reference densities for the multi-electron elements $Z=2$--$118$ with the relativistic B-spline Dirac--Fock code \texttt{atomx} \cite{Zhang2026atomx}, and construct the hydrogen reference from the exact relativistic one-electron Dirac $1s$ solution (Section~\ref{sec:refdata}). The multi-electron calculations use a finite nucleus and a spherical average of open-shell occupations; all 118 records then pass through a common conditioning and transform pipeline. The resulting densities provide a consistent reference for the present fit and remove the need for the Kato cusp.

Building on the earlier parameterisation, we present an element-adaptive parameterisation that addresses all three limitations: (i) the number of basis functions $n_t(Z)$ is selected per element to match the complexity of each atom's electronic structure;
(ii) the cost function simultaneously fits the electron density $\rho(r)$ together with $f_e(g)$ and $f_x(g)$ over an extended range to \SI{36}{\per\angstrom}, thereby enforcing consistency between real-space and reciprocal-space quantities and constraining the high-$g$ tail directly with data; and
(iii) the Kato cusp condition is replaced by an exact $\langle r^4\rangle$ moment constraint, which is physically appropriate for relativistic densities and fixes the $g^4$ coefficient in the expansion of $f_x(g)$ about $g=0$. A fourth advance follows from the demands of a real-space fit in that a non-relativistic basis term is exchanged for a charge-carrying Dirac--Pad\'e term that adds a polynomial-times-exponential shape channel without assigning the fitted term to a physical shell. Importantly, the closed-form analytic framework is retained, so all analytic expressions for the derived quantities remain valid.

It is helpful to examine why the accuracy of the IAM parameterisation matters even though the IAM neglects bonding effects. Bonding modifies the electron density primarily in the valence region at large radial distances, corresponding to the low-$g$ region ($g \lesssim \SI{0.5}{\per\angstrom}$), where structure factors deviate from the IAM \cite{Mahmoudi2025,KappaRefinement2024}. In the higher-$g$ region ($g \gtrsim \SI{2}{\per\angstrom}$), the signal is increasingly dominated by tightly bound core electrons, so the spherical neutral-atom baseline is treated accurately by the IAM and the numerical quality of that baseline becomes important. However, the two regimes involve different error sources and require different remedies: aspherical models such as the transferable aspherical atom model (TAAM) or Hirshfeld atom refinement address the bonding region \cite{KappaRefinement2024}, while the present work targets the spherical neutral-atom contribution at $g\gtrsim\SI{2}{\per\angstrom}$ and its near-nuclear real-space counterpart. Furthermore, experimental methods that measure bonding effects, whether by quantitative CBED \cite{Nakashima2011}, electron ptychography \cite{Hofer2025}, or 3D electron diffraction \cite{Mahmoudi2025}, use an IAM reference as the baseline. Hence, a more accurate baseline reduces one systematic error contribution in any interpretation of the bonding signal.

The improved parameterisation is relevant to a number of modern electron microscopy techniques. High-angle annular dark-field STEM (HAADF-STEM) has been placed on an absolute intensity scale using frozen-phonon multislice \cite{LeBeau2008,LobatoMULTEM2015,VanDyck2009}, which incorporates the undamped high-$g$ electron scattering factor directly and is correspondingly sensitive to its accuracy. Electron ptychography \cite{Jiang2018} and 4D-STEM now achieve sub-\si{\angstrom} resolution, placing stringent demands on the accuracy of the electrostatic potential. Relativistic effects alter the high-angle elastic cross section by about \SI{22}{\percent} for Au beyond \SI{250}{\milli\radian} at \SI{200}{\kilo\electronvolt} \cite{Lentzen2019}. Finally, the parameterised electrostatic potential $V(r)$ can serve as input for computing the Mott elastic cross sections used in backscattered electron (BSE) and electron backscatter diffraction (EBSD) simulations \cite{Salvat2005}, extending the applicability of the parameterisation.

\section{Theory}
\label{sec:theory}

We begin by establishing the notation and recalling the key relationships between scattering factors, the electron density, and the electrostatic potential. We then use the exact relativistic hydrogenic target and its non-relativistic limit to define the rational family from which the Dirac--Pad\'e kernel is constructed, and generalise the resulting basis to multi-electron atoms. Finally, we describe the element-adaptive basis size, exact constraints, working equations, extended cost function, and optimisation procedure.

\subsection{Definitions and notation}
\label{sec:definitions}

Lengths and electron scattering factors are reported in \si{\angstrom}, and reciprocal lengths in \si{\per\angstrom}, following the established scattering-factor convention \cite{DoyleTurner1968,Peng1996,Kirkland1998,Lobato2014}; explicitly identified nuclear dimensions are instead reported in \si{\femto\metre}. The reciprocal-space vector $\mathbf{g}$ is included in the Fourier transform through the kernel $\exp(-2\pi\mathrm{i}\,\mathbf{g}\cdot\mathbf{r})$, and $g=\lvert\mathbf{g}\rvert$ denotes a magnitude throughout this work.

The X-ray scattering factor for a spherically symmetric electron density $\rho(r)$ is defined as
\begin{equation}
f_x(g) = 4\pi \int_0^\infty r^2 \rho(r)\,\frac{\sin(2\pi g r)}{2\pi g r}\,\mathrm{d}r,
\label{eq:fxg}
\end{equation}
where $r$ and $g$ are the magnitudes of the three-dimensional real-space and reciprocal-space vectors $\mathbf{r}$ and $\mathbf{g}$, respectively. The quantity $f_x(g)$ is a dimensionless function that represents the distribution of electrons in reciprocal space. Hence, setting $g = 0$ in equation~\eqref{eq:fxg} gives the charge-neutrality condition for neutral atoms,
\begin{equation}
f_x(0) = \int_0^\infty 4\pi r^2 \rho(r)\,\mathrm{d}r = Z,
\label{eq:neutral}
\end{equation}
with $Z$ the atomic number.

The electron scattering factor $f_e(g)$ is related to $f_x(g)$ \textit{via} the Mott--Bethe formula as \cite{Bethe1930,Mott1930}:
\begin{equation}
f_e(g) = \frac{1}{2\pi^2 a_0}\left[\frac{Z - f_x(g)}{g^2}\right],
\label{eq:feg}
\end{equation}
where $a_0$ is the Bohr radius and $f_e(g)$ has units of \si{\angstrom}. Equation~\eqref{eq:feg} has a removable singularity at $g = 0$; expanding equation~\eqref{eq:fxg} to order $g^2$ about $g = 0$ and inserting the result into equation~\eqref{eq:feg} gives the exact limit derived by \citeasnoun{Ibers1958},
\begin{equation}
f_e(0) = \frac{Z\langle r^2\rangle}{3a_0},
\label{eq:fe0}
\end{equation}
where $\langle r^2\rangle$ is the mean-square radius of the electron distribution, defined by
\begin{equation}
\langle r^2\rangle = \frac{\int_0^\infty r^2 \left[4\pi r^2 \rho(r)\right]\mathrm{d}r}{\int_0^\infty \left[4\pi r^2 \rho(r)\right]\mathrm{d}r}.
\label{eq:r2}
\end{equation}
The electron scattering factor is also related to the electrostatic potential $V(\mathbf{r})$ through its three-dimensional Fourier transform:
\begin{equation}
f_e(\mathbf{g}) = \kappa \int V(\mathbf{r})\exp(-2\pi\mathrm{i}\,\mathbf{g}\cdot\mathbf{r})\,\mathrm{d}\mathbf{r},
\label{eq:feg_V}
\end{equation}
where $\kappa = (4\pi\varepsilon_0)/(2\pi a_0 e)$, with $\varepsilon_0$ the vacuum permittivity and the potential $V$ expressed in volts. For the spherically symmetric case, equation~\eqref{eq:feg_V} reduces to
\begin{equation}
f_e(g) = \kappa \int_0^\infty 4\pi r^2\, V(r)\,\frac{\sin(2\pi g r)}{2\pi g r}\,\mathrm{d}r.
\label{eq:feg_V_sph}
\end{equation}

\subsection{The hydrogenic atom: exact target and Dirac--Pad\'e kernel}
\label{sec:basis}

The definitions above hold for any spherical electron density and leave the form of $\rho(r)$ unspecified. The densities to which the parameterisation is fitted are all-electron relativistic Dirac--Fock calculations (Section~\ref{sec:refdata}), and thus the basis needs a controllable core shape beyond a sum of pure exponentials. The only atom for which the corresponding relativistic shape problem is solvable exactly is the hydrogenic atom: a single electron bound to a point nucleus of charge $Z$. This provides the analytical model used for the basis extension.

\paragraph{Exact relativistic target}
The relativistic ground state is the $1s$ solution of the Dirac equation, whose electron density is known in closed form \cite{Koval2003}:
\begin{equation}
\rho_{1s}^{\mathrm{D}}(r) = \frac{\zeta^{2\gamma+1}}{4\pi\,\Gamma(2\gamma+1)}\,r^{2\gamma-2}\,e^{-\zeta r},
\label{eq:dirac1s}
\end{equation}
with $\gamma = \sqrt{1-(Z\alpha)^2}$, $\zeta = 2Z/a_0$, $\alpha$ the fine-structure constant and $\Gamma$ the gamma function. For the point-Coulomb Dirac $1s$ state, the exponential decay rate $\zeta$ is the same as in the Schr\"odinger $1s$ state at the same $Z$, and the relativistic contraction is carried entirely by the non-integer prefactor $r^{2\gamma-2}$. Writing $b = 4\pi^2/\zeta^2$, its X-ray scattering factor is
\begin{equation}
f_{x,1s}^{\mathrm{D}}(g) =
\frac{\sin\!\left[2\gamma\arctan\!\left(g\sqrt{b}\right)\right]}
{2\gamma g\sqrt{b}\,\left(1+b g^2\right)^\gamma},
\label{eq:fx_dirac1s}
\end{equation}
with the $g=0$ value taken by continuity. The non-integer power gives an algebraic branch point at $g^2=-1/b$. The electrostatic potential of the exact Dirac density involves incomplete gamma functions, and the projected potential $V(R)$ has no finite Bessel-$K$ form of comparable simplicity. The exact relativistic density is therefore the target of the construction rather than its building block.

\paragraph{Non-relativistic limit}
In the non-relativistic limit $\gamma \to 1$, equation~\eqref{eq:dirac1s} reduces exactly to the Schr\"odinger $1s$ density, a pure decaying exponential with the same rate $\zeta$,
\begin{equation}
\rho_{1s}^{\mathrm{NR}}(r) = \frac{\zeta^{3}}{8\pi}\,e^{-\zeta r},
\label{eq:nr1s}
\end{equation}
and every derived quantity becomes elementary \cite{Lobato2014}. The Fourier transform~\eqref{eq:fxg} of this density, which integrates to a single electron, is the rational X-ray scattering factor
\begin{equation}
f_x^{\mathrm{NR}}(g) = \frac{1}{(1+b g^2)^{2}},
\qquad b = \frac{4\pi^{2}}{\zeta^{2}},
\label{eq:fx_nr1s}
\end{equation}
For the neutral hydrogen atom, $Z=1$. Keeping the width explicit rather than substituting the physical value immediately, the Mott--Bethe formula~\eqref{eq:feg} gives the electron-scattering kernel adopted by \citeasnoun{Lobato2014} as the basis function,
\begin{equation}
f_e^{\mathrm{H}}(g;b) =
\frac{b}{2\pi^2a_0}\left[
\frac{1}{1+bg^2}
+ \frac{1}{(1+bg^2)^2}
\right].
\label{eq:feH}
\end{equation}
The physical hydrogen value follows by setting $b=b_{\mathrm{H}}=\pi^2a_0^2$, for which $b_{\mathrm{H}}/(2\pi^2a_0)=a_0/2$. The Lorentzian-plus-square kernel in equation~\eqref{eq:feH}, promoted to a free width and a free amplitude, is a member of the rational family that the parameterisation builds on. The corresponding electrostatic potential is a Yukawa-plus-exponential form with the Yukawa part carrying the correct $1/r$ Coulomb behaviour at the point nucleus. The projected potential $V(R)$, the integral of $V(r)$ along the beam direction with $R$ the two-dimensional radial coordinate perpendicular to it, is therefore a combination of the modified Bessel functions $K_0$ and $K_1$ \cite{Lobato2014}. The non-relativistic limit loses precisely the shape factor $r^{2\gamma-2}$ and the limiting density is a pure exponential with no inflection, whereas the exact density of equation~\eqref{eq:dirac1s} builds up towards the nucleus. A practical approach is therefore not to replace the hydrogenic framework by the exact Dirac function, but to extend the rational family to which equations~\eqref{eq:fx_nr1s} and~\eqref{eq:feH} belong, and to approximate the exact density by a linear combination of its members, each with its own linear amplitude and width.

\paragraph{Rational hierarchy}
Equations~\eqref{eq:fx_nr1s} and~\eqref{eq:feH} are built from integer powers of a single Lorentzian, so the free-width family is organised by the kernels
\begin{equation}
\Phi_n(b,g) = \frac{1}{(1+b g^2)^n}, \qquad n = 1, 2, 3, \dots,
\label{eq:phi_n}
\end{equation}
which form a Pad\'e-type rational hierarchy in $g^2$. These kernels are tied together by one identity that carries the whole construction. Under the Mott--Bethe relation~\eqref{eq:feg}, a charge-carrying Pad\'e power $\Phi_n$ in $f_x(g)$---one that is nonzero at $g = 0$ and therefore carries electron charge---corresponds to the cumulative sum $\Phi_1+\dots+\Phi_n$ in $f_e(g)$ through the telescoping identity
\begin{equation}
g^2\sum_{k=1}^{n}\Phi_k = \frac{1-\Phi_n}{b}.
\label{eq:telescope}
\end{equation}
Taken at $n=2$, this recovers the non-relativistic forms derived above. The term $f_x^{\mathrm{NR}}$ in equation~\eqref{eq:fx_nr1s} is $\Phi_2$, and the corresponding electron scattering factor is the cumulative pair
\begin{equation}
\Phi_1 + \Phi_2 = \frac{(1+bg^2)+1}{(1+bg^2)^2} = \frac{2+bg^2}{(1+bg^2)^2},
\label{eq:nr_n2_member}
\end{equation}
which is exactly the Lorentzian-plus-square kernel of equation~\eqref{eq:feH}. Increasing $n$ within this rational hierarchy gives progressively sharper polynomial-times-exponential density kernels while staying in the pole-only closed-form family. Advancing the hierarchy from $n=2$ to $n=3$ therefore adds the first polynomial-times-exponential shape channel beyond the pure exponential without adopting the non-integer Dirac radial power.

\paragraph{The Dirac--Pad\'e kernel}
The $n=3$ entry is the charge-carrying \emph{Dirac--Pad\'e term} whose name indicates that the term is defined by the hydrogenic Dirac $1s$ density and by the Pad\'e hierarchy above, not that it is assigned to a physical shell in a fitted atom. We denote this member as DP in subsequent expressions. As for every member of the family, this is included in the approximation with its own linear amplitude $a_{\mathrm{DP}}$ (in units of \AA) and width $b_{\mathrm{DP}} > 0$ (in units of \AA$^2$). Its corresponding X-ray scattering factor is the single Pad\'e power
\begin{equation}
f_x^{\mathrm{DP}}(g) = \tilde a_{\mathrm{DP}}\,\Phi_3(b_{\mathrm{DP}},g)
= \frac{2\pi^2 a_0}{b_{\mathrm{DP}}}\,\frac{a_{\mathrm{DP}}}{(1+b_{\mathrm{DP}}g^2)^3},
\label{eq:fx_dp}
\end{equation}
with $\tilde a_{\mathrm{DP}} = 2\pi^2 a_0\,a_{\mathrm{DP}}/b_{\mathrm{DP}}$. The telescoping identity~\eqref{eq:telescope} taken at $n=3$ gives the electron scattering factor in closed form:
\begin{equation}
f_e^{\mathrm{DP}}(g) = a_{\mathrm{DP}}\,\frac{3 + 3\,b_{\mathrm{DP}}g^2 + (b_{\mathrm{DP}}g^2)^2}{(1+b_{\mathrm{DP}}g^2)^3}, \label{eq:fe_dp}
\end{equation}
with a corresponding electron-density kernel
\begin{equation}
\rho_{\mathrm{DP}}(r) = \frac{\hat a_{\mathrm{DP}}}{4}\bigl(1 + b'_{\mathrm{DP}} r\bigr)\,e^{-b'_{\mathrm{DP}} r},
\qquad b'_{\mathrm{DP}} = \frac{2\pi}{b_{\mathrm{DP}}^{1/2}}.
\label{eq:rho_dp}
\end{equation}
The amplitude is $\hat a_{\mathrm{DP}} = 2\pi^4 a_0\,a_{\mathrm{DP}}/b_{\mathrm{DP}}^{5/2}$, and the $(1+b'_{\mathrm{DP}} r)$ factor is the important difference from the non-relativistic $1s$ kernel. This introduces one inflection in the density kernel $\rho_{\mathrm{DP}}(r)$, and the kernel multiplying $\hat a_{\mathrm{DP}}$ is non-negative for $r \ge 0$. This is why the Pad\'e-type construction is carried out in $f_x(g)$ rather than in $f_e(g)$. For $n\geq2$, a unit positive $\Phi_n$ contribution in $f_x$ inverts to a density kernel
\begin{equation}
\rho^{(n)}(r) \propto P_{n-2}(b' r)\,e^{-b' r},
\label{eq:pade_inversion}
\end{equation}
where $P_{n-2}$ is a polynomial with non-negative integer coefficients,
\begin{equation}
P_0 = 1, \qquad P_1 = 1+b' r, \qquad P_2 = 3+3b' r + (b' r)^2,
\label{eq:pade_polys}
\end{equation}
so the kernel itself is non-negative for every $r$. The density remains a polynomial times an exponential, so every derived quantity remains in closed form, with $V(r)$ again a sum of Yukawa-plus-exponential terms. The projected potential $V(R)$ follows from the same Bessel-$K$ line-projection identities, now involving $K_0$, $K_1$ and $K_2$ in place of $K_0$ and $K_1$.

\paragraph{Correction directly in $f_e(g)$}
The same hierarchy might suggest a simpler route: adding the sharper Pad\'e power directly to $f_e(g)$. However, this route fails two requirements of the present basis. First, a term that replaces one charge-carrying member of the hydrogenic expansion must retain the Mott--Bethe high-$g$ tail of $f_e(g)$. The Dirac--Pad\'e member decays as $1/g^2$ in $f_e(g)$, whereas an isolated $\Phi_3$ term inserted directly in $f_e(g)$ decays as $1/g^6$ for $g\to\infty$. Hence, while the former remains compatible with a charge-carrying expansion term, the latter has the wrong asymptotic decay. Second, the associated density shape should not contain a built-in sign change. Signed amplitudes are essential in the final expansion, but a signed amplitude should add or subtract a complete radial shape and should not force a positive--negative redistribution within one basis member. The Dirac--Pad\'e construction satisfies both requirements: its $1/g^2$ high-$g$ tail preserves the charge-carrying role, while the $f_x^{\mathrm{DP}}(g)\propto\Phi_3$ definition in equation~\eqref{eq:fx_dp} gives the non-negative density kernel of equation~\eqref{eq:rho_dp}. An isolated additive correction $\Delta f_e(g)=a\,\Phi_3(b,g)$ instead gives, through the Mott--Bethe relation,
\begin{align}
\Delta f_x(g)
&= -2\pi^2 a_0\,g^2\Delta f_e(g) \nonumber\\
&= -\frac{2\pi^2 a_0 a}{b}\bigl[\Phi_2(b,g)-\Phi_3(b,g)\bigr],
\label{eq:bad_fe_pade}
\end{align}
where the difference of equation~\eqref{eq:telescope} between $n=3$ and $n=2$, $g^2\Phi_3=(\Phi_2-\Phi_3)/b$, has been used. Equation~\eqref{eq:bad_fe_pade} gives $\Delta f_x(0)=0$, and thus the associated density contribution has zero volume integral. The inverse transform makes this zero-charge redistribution explicit:
\begin{equation}
\Delta\rho(r) \propto \bigl(3-b' r\bigr)e^{-b' r},
\qquad b'=\frac{2\pi}{b^{1/2}},
\label{eq:bad_fe_density}
\end{equation}
which changes sign at $r=3/b'$. Depending on the sign of $a$, the same scalar coefficient must add density near the nucleus and remove it outside, or the reverse. Such a term may be useful for a deliberately zero-charge redistribution, but it is the wrong function for replacing one charge-carrying member of the hydrogenic expansion. The sign change also propagates to the associated electrostatic-potential contribution. The Pad\'e-type extension is therefore introduced in $f_x(g)$, where the charge-carrying density kernel is non-negative, rather than as an unconstrained additive Pad\'e correction in $f_e(g)$.

\begin{figure}[t]
\centering
\includegraphics[width=\textwidth]{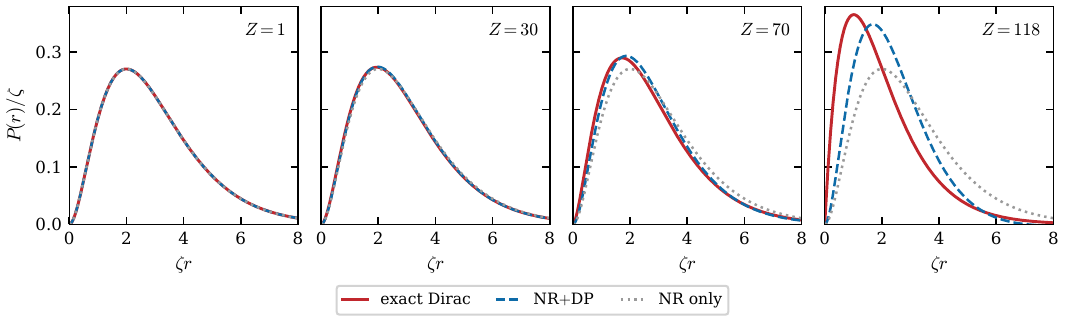}
\caption{Exact point-Coulomb Dirac $1s$ radial probability $4\pi r^2\rho(r)$ (solid), two-term NR$+$DP moment-matched combination of one non-relativistic and one Dirac--Pad\'e term (dashed), and a single charge-matched non-relativistic term (dotted), plotted against the dimensionless radius $\zeta r$ for four nuclear charges. All three curves share the decay $\zeta=2Z/a_0$, so the comparison isolates the relativistic inner shape provided by the factor $r^{2\gamma-2}$ and approximated by the Dirac--Pad\'e term.}
\label{fig:dirac}
\end{figure}

\paragraph{Comparison with the exact relativistic density}
The construction can now be tested against the exact relativistic target of equation~\eqref{eq:dirac1s}, which is defined in the subcritical point-Coulomb domain $Z\alpha<1$. Matching the charge and $\langle r^2\rangle$ of $\rho_{1s}^{\mathrm{D}}$ with one non-relativistic and one Dirac--Pad\'e term fixes both amplitudes and yields the two-term (NR$+$DP) moment-matched approximation
\begin{equation}
\rho_{\mathrm{NR}+\mathrm{DP}}(r) = A(\gamma)\,\frac{\zeta^3}{8\pi}\,e^{-\zeta r}
+ B(\gamma)\,\frac{\zeta^3}{32\pi}\,(1+\zeta r)\,e^{-\zeta r},
\label{eq:pade03}
\end{equation}
with $A(\gamma) = (8-3\gamma-2\gamma^2)/3$ and $B(\gamma) = (2\gamma^2+3\gamma-5)/3$. Here, the first kernel is the non-relativistic density of equation~\eqref{eq:nr1s} and the second is the Dirac--Pad\'e density kernel of equation~\eqref{eq:rho_dp} at $b'_{\mathrm{DP}} = \zeta$, so $A(\gamma)$ and $B(\gamma)$ are the linear amplitudes of the family approximation at the hydrogenic level, fixed by the two moments rather than by fitting. Figure~\ref{fig:dirac} compares the radial probability $4\pi r^2\rho_{\mathrm{NR}+\mathrm{DP}}(r)$ with the exact Dirac probability $4\pi r^2\rho_{1s}^{\mathrm{D}}(r)$ and with a single charge-matched non-relativistic term of unit amplitude with the same decay $\zeta$. Sharing the point-Coulomb decay across all three curves isolates the relativistic inner shape generated by $r^{2\gamma-2}$. At $Z = 1$, the three nearly coincide and $B$ is negligible (${\sim}{-}6\times10^{-5}$; exactly zero only in the non-relativistic limit $\gamma = 1$); with increasing $Z$, the single non-relativistic term peaks too broadly and does not reproduce the build-up of density towards the nucleus, while the two-term NR$+$DP combination, matched in charge and $\langle r^2\rangle$, follows the exact density throughout the present reference range. Its endpoint at $Z=118$ is set by dataset coverage, not by the analytic comparison; the residual there is the leading error of the two-term form, in which the $r^{2\gamma-2}$ cusp lies outside the polynomial-times-exponential family.

\subsection{Generalisation to multi-electron atoms}
\label{sec:dirac_pade}

Following the earlier construction \cite{Lobato2014}, the single-electron forms of Section~\ref{sec:basis} are now generalised to a multi-electron atom, whose shells span widely different length scales. We relax the fixed width $b = 4\pi^2/\zeta^2$ in equation~\eqref{eq:fx_nr1s} to obtain a free width for each term, and use $M$ non-relativistic terms with signed amplitudes to represent the shell structure. At the hydrogenic level, relativistic contraction changes the inner density shape away from a pure exponential. We therefore exchange one non-relativistic term for a single $n=3$ Pad\'e member, giving the constrained expansion a polynomial-times-exponential shape channel that the pure non-relativistic basis lacks. The Dirac--Pad\'e production basis therefore expands the electron scattering factor in $M$ non-relativistic members of the rational family plus one Dirac--Pad\'e term:
\begin{align}
f_e(g) &= \sum_{i=1}^{M} a_i\,\frac{2+b_ig^2}{(1+b_ig^2)^2} \notag\\
&\quad + a_{\mathrm{DP}}\,\frac{3 + 3\,b_{\mathrm{DP}}g^2 + (b_{\mathrm{DP}}g^2)^2}{(1+b_{\mathrm{DP}}g^2)^3},
\label{eq:feg_param}
\end{align}
where $a_i$ and $a_{\mathrm{DP}}$ are linear coefficients (in units of \AA) and $b_i, b_{\mathrm{DP}} > 0$ are nonlinear width parameters (in units of \AA$^2$). Each non-relativistic term is the Lorentzian-plus-square kernel of equation~\eqref{eq:nr_n2_member}, with an electron density $\rho_i(r) \propto \exp\bigl(-2\pi r/b_i^{1/2}\bigr)$: the exponential of equation~\eqref{eq:nr1s} with a generalised decay. The final term is the Dirac--Pad\'e kernel in equations~\eqref{eq:fx_dp}--\eqref{eq:rho_dp}. The crucial difference from that construction is that the total number of terms $n_t = M + 1$ is now an element-dependent quantity, $n_t = n_t(Z)$, assigned as described in the shell-filling schedule of Section~\ref{sec:adaptive}, rather than being fixed at five for all atoms. Substituting equation~\eqref{eq:feg_param} into the Mott--Bethe relation~\eqref{eq:feg} and solving for $f_x(g)$ yields the raw X-ray scattering factor:
\begin{equation}
\begin{aligned}
f_x(g) = Z - 2\pi^2 a_0 g^2 \Biggl[\, &\sum_{i=1}^{M} a_i\,\frac{2+b_ig^2}{(1+b_ig^2)^2} \\
&+ a_{\mathrm{DP}}\,\frac{3 + 3\,b_{\mathrm{DP}}g^2 + (b_{\mathrm{DP}}g^2)^2}{(1+b_{\mathrm{DP}}g^2)^3} \Biggr].
\end{aligned}
\label{eq:fxg_param}
\end{equation}
All derived quantities can be obtained in closed form, term by term (Section~\ref{sec:basis}) using the explicit expressions for $V(r)$, $V(R)$ and the electron density $\rho(r)$, the latter additionally requiring the charge-neutrality constraint of Section~\ref{sec:constraints}. These are summarised in the working equations of Section~\ref{sec:working_eqs}.

With $M = n_t - 1$, the Dirac--Pad\'e basis carries exactly the same $2n_t$ parameters as a pure non-relativistic expansion with $n_t$ terms. Importantly, all amplitudes, including $a_{\mathrm{DP}}$, remain signed expansion coefficients. The term name does not imply that one fitted basis function is assigned to a specific physical shell of a multi-electron atom, and the Dirac--Pad\'e term should not be interpreted as a physical decomposition of the relativistic excess. The equal-count non-relativistic baseline is an alternative model with $n_t$ non-relativistic terms and no Dirac--Pad\'e term, rather than the Dirac--Pad\'e model with $a_{\mathrm{DP}}$ set to zero. We refer to this enriched basis as ``Dirac--Pad\'e'' and to the equal-count baseline as the ``matched NR'' basis. Section~\ref{sec:res_dirac_pade} quantifies the difference between them.

\subsection{Element-adaptive basis size}
\label{sec:adaptive}

The generalisation of Section~\ref{sec:dirac_pade} leaves the number of terms required for each element undetermined. The number of basis functions $n_t(Z)$ is assigned phenomenologically from the fixed shell-filling schedule in Table~\ref{tab:nt}. A newly occupied principal shell introduces an additional peak or shoulder and the adjacent curvature in $4\pi r^2\rho(r)$; the schedule therefore adds two basis terms at each noble-gas boundary beyond helium (and a single term at helium itself). This provides the signed expansion with the flexibility needed to represent the new radial feature and its surrounding slope while keeping the light-element basis compact.

A single fixed basis size cannot achieve this uniformly across the periodic table: the number of occupied principal shells grows from one in hydrogen to seven in period 7, and their characteristic radii span two to three orders of magnitude---from the tightly bound K shell, whose hydrogenic length scale contracts as $1/Z$, out to diffuse valence shells of order \SI{1}{\angstrom}. Five terms suffice through Ne in the adopted schedule, whereas Na ($Z=11$) is the first element assigned more than five terms; the basis size subsequently increases at each new principal-shell period (Table~\ref{tab:nt}). A fixed small basis must then either over-parameterise the light elements or leave later-period elements with too little flexibility to reproduce their shell-resolved radial distribution, so the relative accuracy degrades steeply with $Z$ (Section~\ref{sec:res_feg}). Increasing $n_t$ with the observed shell complexity instead keeps the relative accuracy approximately uniform. Too few terms underfit, whereas unnecessary terms can create near-degenerate widths and drive the constrained amplitude solve towards ill-conditioning (Section~\ref{sec:optim}).

Hydrogen takes the smallest total basis, $n_t = 2$. Its $Z=1$ reference is constructed from the exact relativistic Dirac $1s$ density of equation~\eqref{eq:dirac1s} and passed through the same conditioning and transform pipeline as the multi-electron reference densities (Section~\ref{sec:refdata}). This target is distinct from the analytic non-relativistic hydrogen forms used as basis functions in Section~\ref{sec:basis}. Hydrogen is therefore fitted under the same procedure as every other element---with two non-relativistic terms in the matched NR basis, or one non-relativistic and one Dirac--Pad\'e term in the Dirac--Pad\'e basis---and satisfies the first two exact constraints of Section~\ref{sec:constraints}.

\begin{table}[t]
\centering
\caption{The element-adaptive basis size $n_t(Z)$ follows a fixed shell-filling schedule, stepping up at the noble-gas boundaries; hydrogen takes the smallest basis ($n_t = 2$). Both bases carry the same total parameter count $2n_t$ (Section~\ref{sec:dirac_pade}).}
\label{tab:nt}
\begin{tabular}{lccc}
\toprule
Row / block                          & $Z$ range       & $n_t$ & params \\
\midrule
H                                    & $1$             & $2$  & $4$  \\
He                                   & $2$             & $3$  & $6$  \\
Period 2 (Li--Ne)                    & $3$--$10$       & $5$  & $10$ \\
Period 3 (Na--Ar)                    & $11$--$18$      & $7$  & $14$ \\
Period 4 (K--Kr)                     & $19$--$36$      & $9$  & $18$ \\
Period 5 (Rb--Xe)                    & $37$--$54$      & $11$ & $22$ \\
Period 6 (Cs--Rn)                    & $55$--$86$      & $13$ & $26$ \\
Period 7 (Fr--Og)                    & $87$--$118$     & $15$ & $30$ \\
\bottomrule
\end{tabular}

\end{table}

\subsection{Exact constraints and analytic density}
\label{sec:constraints}

The amplitudes in the Dirac--Pad\'e expansion of equation~\eqref{eq:feg_param} are signed coefficients, and nothing so far ties the summed density to the exact integral properties of the reference atom. Three conditions do exactly that and are imposed analytically rather than fitted: charge neutrality, the mean-square radius $\langle r^2\rangle$ through $f_e(0)$, and the fourth radial moment $\langle r^4\rangle$.

\paragraph{Charge neutrality from Riemann--Lebesgue} The parameterised density must be integrable (its volume integral equals $Z$), so by the Riemann--Lebesgue lemma the X-ray scattering factor must vanish at infinity, $f_x(g)\to 0$ as $g\to\infty$. Evaluating the raw scattering factor of equation~\eqref{eq:fxg_param} at large $g$ with the per-basis-function limits
\begin{equation}
\begin{aligned}
g^2\,\frac{2+b g^2}{(1+b g^2)^2} &\to \frac{1}{b}, \\[2pt]
g^2\,\frac{3+3b_{\mathrm{DP}}g^2+(b_{\mathrm{DP}}g^2)^2}{(1+b_{\mathrm{DP}}g^2)^3} &\to \frac{1}{b_{\mathrm{DP}}} \quad (g\to\infty),
\end{aligned}
\label{eq:largeg_limits}
\end{equation}
both following directly from the telescoping identity~\eqref{eq:telescope}, whose right-hand side tends to $1/b$ as $g\to\infty$. Weighting each limit by its amplitude gives $f_x(g\to\infty) = Z - 2\pi^2 a_0\bigl(\sum_{i} a_i/b_i + a_{\mathrm{DP}}/b_{\mathrm{DP}}\bigr)$, and setting this to zero yields the first exact linear constraint,
\begin{equation}
\sum_{i=1}^{M} \frac{a_i}{b_i} \;+\; \frac{a_{\mathrm{DP}}}{b_{\mathrm{DP}}} \;=\; \frac{Z}{2\pi^2\,a_0}.
\label{eq:cstr_charge}
\end{equation}
Here and in the Dirac--Pad\'e closed forms below the sum runs over the $M = n_t-1$ non-relativistic terms and the final term is the Dirac--Pad\'e contribution. For the matched-NR basis the Dirac--Pad\'e column is omitted and the non-relativistic sum instead runs over all $n_t$ terms (Section~\ref{sec:dirac_pade}).

\paragraph{Analytic electron density} Applying the charge-neutrality constraint~\eqref{eq:cstr_charge} cancels the constant $Z$ in the raw X-ray scattering factor and reduces it to the compact form
\begin{equation}
f_x(g) = 2\pi^2 a_0 \left[ \sum_{i=1}^{M} \frac{a_i}{b_i(1+b_ig^2)^2}
\;+\; \frac{a_{\mathrm{DP}}}{b_{\mathrm{DP}}(1+b_{\mathrm{DP}}g^2)^3} \right],
\label{eq:fx_compact}
\end{equation}
in which the sum collects the $M$ non-relativistic terms and the final term is the Dirac--Pad\'e contribution $f_x^{\mathrm{DP}}$ of equation~\eqref{eq:fx_dp}. Its inverse three-dimensional Fourier transform gives the electron density in the matching closed form,
\begin{equation}
\begin{aligned}
\rho(r) = 2\pi^4 a_0 \Bigg[ &\sum_{i=1}^{M} \frac{a_i}{b_i^{5/2}}\, e^{-2\pi r/b_i^{1/2}} \\
&+\; \frac{a_{\mathrm{DP}}}{4\,b_{\mathrm{DP}}^{5/2}}\Bigl(1 + \tfrac{2\pi r}{b_{\mathrm{DP}}^{1/2}}\Bigr)\, e^{-2\pi r/b_{\mathrm{DP}}^{1/2}} \Bigg],
\end{aligned}
\label{eq:rhor}
\end{equation}
a sum of near-exponential functions, consistent with the observation that atomic electron densities are nearly piecewise exponential \cite{Sperber1971,Wang1977}. The final term is the Dirac--Pad\'e density of equation~\eqref{eq:rho_dp}, written here in full; it carries the same $b_{\mathrm{DP}}^{-5/2}$ width scaling as a non-relativistic term, up to the $1/4$ prefactor and the $(1+b'_{\mathrm{DP}} r)$ dressing.

\paragraph{Radial moments and exact constraints} For a non-negative integer $n$, define the radial moment
\begin{equation}
p_n \equiv 4\pi\int_0^\infty r^{n+2}\rho(r)\,\mathrm{d}r = Z\langle r^n\rangle.
\label{eq:moment_def}
\end{equation}
Substituting equation~\eqref{eq:rhor} gives the closed form
\begin{equation}
\begin{split}
p_n \;=\; 2^{-n}\,\pi^{\,2-n}\,a_0\,(n+2)! \Bigg[ &\sum_{i=1}^{M} a_i\,b_i^{(n-2)/2} \\
&+\; \frac{n+4}{4}\,a_{\mathrm{DP}}\,b_{\mathrm{DP}}^{(n-2)/2} \Bigg],
\end{split}
\label{eq:moment_general}
\end{equation}
so each moment is an exact linear function of the amplitudes at fixed widths $\{b_i, b_{\mathrm{DP}}\}$. Thus, equation~\eqref{eq:moment_general} applies to radial moments of any non-negative integer order. To identify which moments occur successively in the expansion of the X-ray scattering factor about $g=0$, expand the sinc kernel of equation~\eqref{eq:fxg}:
\begin{equation}
f_x(g) \;=\; \sum_{j=0}^{\infty} \frac{(-1)^j\,(2\pi g)^{2j}}{(2j+1)!}\,p_{2j}.
\label{eq:fxg_moments}
\end{equation}
Only the even moments $p_0,p_2,p_4,\ldots$ occur as Taylor coefficients in equation~\eqref{eq:fxg_moments}. The first three therefore provide the successive low-$g$ constraints used here. For $n=0,2,4$, the Dirac--Pad\'e factor $(n+4)/4$, which arises from the $(1+b'_{\mathrm{DP}} r)$ dressing of $\rho_{\mathrm{DP}}$, equals $1$, $\tfrac{3}{2}$ and $2$, respectively. The complete continuous transform determines $\rho(r)$ and therefore all of its moments in principle, but an odd radial moment is not an independent Taylor coefficient at $g=0$. The selected higher even moments control successive even derivatives of $f_x(g)$ at $g=0$ but do not by themselves constrain its asymptotic high-$g$ accuracy. The $n=0$ case is exceptional: for this basis, charge normalisation $p_0=Z$ is algebraically identical to the cancellation condition $f_x(g\to\infty)=0$ in equation~\eqref{eq:cstr_charge}. The $n=2$ case, with $p_2 = Z\langle r^2\rangle$, recovers the Ibers relation~\eqref{eq:fe0}, which in terms of the amplitudes reads
\begin{equation}
\sum_{i=1}^{M} 2a_i \;+\; 3a_{\mathrm{DP}} \;=\; f_e(0),
\label{eq:cstr_r2}
\end{equation}
the Dirac--Pad\'e coefficient $3 = 2\times\tfrac{3}{2}$ being the $(n+4)/4$ factor of equation~\eqref{eq:moment_general} at $n=2$ (equivalently, the numerator of equation~\eqref{eq:fe_dp} at $g=0$). The $n=4$ case pins $\langle r^4\rangle$,
\begin{equation}
\sum_{i=1}^{M} a_i b_i \;+\; 2a_{\mathrm{DP}} b_{\mathrm{DP}} \;=\; \frac{\pi^2 p_4}{45 a_0},
\label{eq:cstr_r4}
\end{equation}
and fixes the $g^4$ coefficient of $f_x(g)$.

\paragraph{Replacing the Kato cusp by $\langle r^4\rangle$} In the earlier parameterisation \cite{Lobato2014}, the third exact constraint was the Kato cusp condition \cite{Kato1957,Steiner1963},
\begin{equation}
\left.\frac{\partial\rho}{\partial r}\right|_{r=0} = -\frac{2Z}{a_0}\,\rho(0),
\label{eq:kato}
\end{equation}
a local condition on the derivative of $\rho$ at the nucleus. This condition is derived for a non-relativistic point-nucleus Hamiltonian and is not satisfied by either type of reference density used here. The point-Coulomb Dirac density for hydrogen behaves as $\rho(r)\propto r^{2\gamma-2}$, with $\gamma = \sqrt{1-(Z\alpha)^2}<1$, whereas the finite-Fermi-nucleus densities for the multi-electron atoms are regular and have zero radial derivative at the origin; neither obeys equation~\eqref{eq:kato} \cite{Mastalerz2010}. Moreover, the representable near-nuclear density for $r\geq r^*$ is included directly in the fit, while the unrepresentable region $r<r^*$ is excluded as described in Section~\ref{sec:cost}. We therefore use the $\langle r^4\rangle$ condition of equation~\eqref{eq:cstr_r4} as the third exact constraint. This retains $n_t-3$ free linear degrees of freedom for atoms with at least three fitted terms; hydrogen has two terms and uses only charge neutrality and $\langle r^2\rangle$.

\paragraph{The constraint linear system} For atoms with at least three fitted terms, the three constraints---charge neutrality~\eqref{eq:cstr_charge}, $\langle r^2\rangle$~\eqref{eq:cstr_r2} and $\langle r^4\rangle$~\eqref{eq:cstr_r4}---are linear in the amplitudes at fixed widths and are imposed exactly for both bases. Hydrogen has only two fitted terms and therefore uses the first two constraints only. For the Dirac--Pad\'e basis, collecting the constraints as the rows of one linear system $T\mathbf{a} = \mathbf{c}$, with amplitude vector $\mathbf{a} = (a_1,\dots,a_M,a_{\mathrm{DP}})^{\mathsf{T}}$, gives the following non-relativistic and Dirac--Pad\'e columns:
\begin{equation}
t_i^{\mathrm{NR}} = \begin{pmatrix} 1/b_i \\ 2 \\ b_i \end{pmatrix}, \qquad
t_{\mathrm{DP}} = \begin{pmatrix} 1/b_{\mathrm{DP}} \\ 3 \\ 2b_{\mathrm{DP}} \end{pmatrix},
\label{eq:cstr_columns}
\end{equation}
with right-hand side $\mathbf{c} = \bigl(Z/2\pi^2 a_0,\; f_e(0),\; \pi^2 p_4/45 a_0\bigr)^{\mathsf{T}}$, so the three rows reproduce equations~\eqref{eq:cstr_charge}, \eqref{eq:cstr_r2} and \eqref{eq:cstr_r4} in order of increasing moment index. The matched-NR basis uses only columns of type $t_i^{\mathrm{NR}}$, one for each of its $n_t$ non-relativistic terms. The implementation divides the second row and its right-hand side by two; this algebraically equivalent scaling gives coefficients $1$ and $3/2$ for the Dirac--Pad\'e basis and right-hand side $f_e(0)/2$, and improves numerical conditioning without changing the constraint. The Dirac--Pad\'e column is the non-relativistic column evaluated at $b_{\mathrm{DP}}$, reweighted row by row by the factor $(n+4)/4$ of equation~\eqref{eq:moment_general}, so the Dirac--Pad\'e amplitude carries charge, mean-square radius and $\langle r^4\rangle$ on exactly the same footing as the non-relativistic terms. For distinct widths the rows of $T$ are independent, so the admissible amplitudes form a particular solution of $T\mathbf{a} = \mathbf{c}$ plus the null space of $T$. At every width trial, the inner least-squares solve of Section~\ref{sec:optim} chooses the amplitudes within this null space; the outer optimiser varies only the nonlinear widths.

\subsection{Working equations for implementation}
\label{sec:working_eqs}

The complete Dirac--Pad\'e parameterisation is collected here in closed form for direct use. Every observable is a sum of $M = n_t - 1$ non-relativistic terms plus one Dirac--Pad\'e term, sharing the same primary parameter set $\{(a_i,b_i)\}_{i=1}^{M}$ and $(a_{\mathrm{DP}},b_{\mathrm{DP}})$. The matched-NR baseline follows from the same non-relativistic forms with the Dirac--Pad\'e contribution omitted and the sum extended to its own $n_t$ non-relativistic terms. Once a model's coefficients have been fitted, the Dirac--Pad\'e observables follow from
\begin{align}
f_e(g) &= \sum_{i=1}^{M} a_i\,\frac{2+b_ig^2}{(1+b_ig^2)^2} \notag\\
&\quad + a_{\mathrm{DP}}\,\frac{3+3b_{\mathrm{DP}}g^2+(b_{\mathrm{DP}}g^2)^2}{(1+b_{\mathrm{DP}}g^2)^3}, \label{eq:work_fe}\\
f_x(g) &= \sum_{i=1}^{M} \frac{\tilde a_i}{(1+b_ig^2)^2}
        + \frac{\tilde a_{\mathrm{DP}}}{(1+b_{\mathrm{DP}}g^2)^3}, \label{eq:work_fx}\\
\rho(r) &= \sum_{i=1}^{M} \hat a_i\,e^{-b'_i r} \notag\\
&\quad + \frac{\hat a_{\mathrm{DP}}}{4}\,(1+b'_{\mathrm{DP}} r)\,e^{-b'_{\mathrm{DP}} r}, \label{eq:work_rho}\\
P(r) &= 4\pi r^2\,\rho(r), \label{eq:work_P}\\
V(r) &= \sum_{i=1}^{M} a'_i\,e^{-b'_i r}\!\left[\frac{2}{b'_i r}+1\right] \notag\\
&\quad + \frac{a'_{\mathrm{DP}}}{4}\,e^{-b'_{\mathrm{DP}} r}
\!\left[\frac{8}{b'_{\mathrm{DP}} r}+5+b'_{\mathrm{DP}} r\right]. \label{eq:work_Vr}
\end{align}
Finally, the projected potential is
\begin{multline}
V(R) = \sum_{i=1}^{M} a''_i\!\left[\frac{2K_0(b'_i R)}{b'_i}+R\,K_1(b'_i R)\right] \\
+ \frac{a''_{\mathrm{DP}}}{8}\Bigl[\Bigl(\frac{16}{b'_{\mathrm{DP}}}+b'_{\mathrm{DP}} R^2\Bigr)K_0(b'_{\mathrm{DP}} R) \\
+ 10\,R\,K_1(b'_{\mathrm{DP}} R) + b'_{\mathrm{DP}} R^2\,K_2(b'_{\mathrm{DP}} R)\Bigr],
\label{eq:work_VR}
\end{multline}
where $K_0$, $K_1$ and $K_2$ are the modified Bessel functions of the second kind, $r$ ($R$) is the three- (two-) dimensional radial coordinate, and the derived per-term coefficients, identical in form for the non-relativistic and Dirac--Pad\'e slots ($t\in\{1,\dots,M,\mathrm{DP}\}$), are
\begin{equation}
\begin{aligned}
&b'_t = \frac{2\pi}{\sqrt{b_t}},\qquad
\tilde a_t = \frac{2\pi^2 a_0\,a_t}{b_t},\qquad
\hat a_t = \frac{2\pi^4 a_0\,a_t}{b_t^{5/2}}, \\[3pt]
&a'_t = \frac{\pi^2 a_t}{\kappa\,b_t^{3/2}},\qquad
a''_t = 2\,a'_t.
\end{aligned}
\label{eq:working_coeffs}
\end{equation}
The fitted linear amplitudes $\{a_t\}$ and nonlinear widths $\{b_t\}$ are the only quantities that need to be tabulated; the derived coefficients $(b'_t,\tilde a_t,\hat a_t,a'_t,a''_t)$ and hence all six observables follow from equation~\eqref{eq:working_coeffs} in closed form. The non-relativistic parts of $V(r)$ and $V(R)$ are those derived previously \cite{Lobato2014}; the Dirac--Pad\'e parts follow in the same way, by solving Poisson's equation for the density kernel of equation~\eqref{eq:rho_dp} and integrating the resulting potential along the beam direction, which is what introduces the $K_2$ contribution.

\subsection{Extended cost function}
\label{sec:cost}

A central fitting advance of this work is the simultaneous fitting of reciprocal-space and real-space quantities. The earlier cost function \cite{Lobato2014} contained only data residuals on $f_e(g)$ and $f_x(g)$ (beyond the exact constraints and positivity); the present cost is the weighted sum
\begin{equation}
\chi^2 \;=\; \sum_i w_i\,\sigma_i
\label{eq:cost}
\end{equation}
of dimensionless terms $\sigma_i$ drawn from four families, each evaluated as a normalised mean absolute deviation or a zero-when-satisfied penalty so that all terms are directly comparable across the periodic table and individual terms can be emphasised or switched off without retuning the optimiser. The components and their weights $w_i$ are collected in Table~\ref{tab:cost}; the tabulated $w_i$ are the effective weights of the fixed production configuration (the per-term weights multiplied by a global scale factor of 2000), referred to below as the canonical weight set and used unchanged for every element. The scalar is written $\chi^2$ by convention although its data channels are normalised mean absolute deviations rather than least-squares residuals.

The first family comprises the \emph{data channels} on $f_e(g)$, $f_x(g)$ and the electron density $\rho(r)$: for each quantity a plain residual, a weighted residual (the $g^2$-weighted forms emphasising the high-$g$ tail of $f_e$ and $f_x$, the $r^2$-weighted form emphasising the outer-shell radial distribution $4\pi r^2\rho$), and a log-space residual, together with a log--log slope residual on $r^2\rho(r)$ that penalises shape errors directly. The second family enforces \emph{physicality}: positivity and monotone-decay penalties on $f_e$, $f_x$, $\rho$ and $V$, which are zero when the constraint is satisfied and non-zero only when the fit violates it. The third family is a set of \emph{characteristic-radii} penalties: six disjoint families of landmark radii of $4\pi r^2\rho(r)$ (peaks, shoulders, valleys, inflections, and rising- and falling-flank tail anchors) are extracted from the reference density and the fitted curve is held to the reference height at each, capturing the localised shell features that a grid-averaged residual under-weights. The fourth family is a small set of \emph{anti-cancellation regularisers} at the curve origins that discourage large opposite-sign coefficient pairs.

The reciprocal-space physicality terms retain a curve-scale normalisation: for $y\in\{f_e,f_x\}$, the positivity penalty is $\langle[-y(g_j)]_+\rangle/\max|y_{\mathrm{ref}}|$, where $[x]_+=\max(x,0)$. The 64-point exponentially spaced constraint grid extends from an origin-adjacent positive value to $1.2g_{\max}$, with $g_{\max}$ the outermost reference sample. By contrast, a single global normalisation is unsuitable for the real-space quantities because their component magnitudes span many orders of magnitude over the atom-specific radial interval $0<r\leq r_{\max}$. A negative tail can therefore give a negligible globally normalised penalty, while direct summation of the exponentially decaying signed components can underflow to zero and conceal its sign. With $\rho_k(r)$ denoting the individual terms of equation~\eqref{eq:work_rho} and $W_k(r)=rV_k(r)$ those of equation~\eqref{eq:work_Vr}, the real-space physicality terms therefore use the local component-normalised ratios
\begin{equation}
q_\rho(r)=\frac{\sum_k\rho_k(r)}{\sum_k|\rho_k(r)|},\qquad
q_W(r)=\frac{\sum_k W_k(r)}{\sum_k|W_k(r)|},\qquad W(r)=rV(r).
\label{eq:realspace_positivity_ratios}
\end{equation}
The common slowest exponential is removed before evaluating each ratio, so it cancels analytically rather than underflowing. The denominator then provides a local $L^1$ component scale, so each ratio retains the relative sign and degree of cancellation independently of the absolute tail amplitude. Because $r>0$ on the soft grid, $W$ and $V$ have the same sign, while $W$ avoids the Coulomb singularity of $V$ at the origin. On 256 logarithmically spaced radii $r_j\in[\SI{0.01}{\angstrom},r_{\max}]$, where $r_{\max}$ is the atom-specific outer radius of the deposited reference grid, the real-space positivity terms are
\begin{equation}
\sigma_{\mathrm{pos}}^{(y)}=
\max\!\left\{\max_j[-q_y(r_j)]_+,\,
\left[\eta_\Sigma-\frac{A_L}{\sum_k|a_k|}\right]_+\right\},
\qquad y\in\{\rho,W\}.
\label{eq:realspace_soft_positivity}
\end{equation}
Here $A_L$ is the effective amplitude of the first non-cancelling, slowest-decaying group after quantisation to the delivery precision, with the higher-degree Dirac--Pad\'e polynomial taking precedence when decay exponents coincide, and $\eta_\Sigma$ is the corresponding summation-roundoff margin. The second term in equation~\eqref{eq:realspace_soft_positivity} supplies a finite optimisation signal when the analytic tail has the wrong sign. Monotone-decay penalties remain normalised mean positive increments of the sampled physical curves. These sampled terms guide the optimisation; final real-space admissibility is decided separately in delivery precision as described in Section~\ref{sec:optim}.

The inclusion of the real-space density removes an ambiguity of fitting a finite, discretely sampled reciprocal-space interval: density differences at unresolved radial scales can produce numerically indistinguishable scattering-factor samples after integration with the oscillating kernel $\sin(2\pi g r)/(2\pi g r)$. Constraining the fit simultaneously in both spaces requires the resulting density to match the reference shell structure rather than merely reproducing the sampled transform. The innermost finite-nucleus plateau is excluded from the direct density residuals by the physical cutoff
\begin{equation}
r^* = 0.40\,r_N(Z)
\label{eq:rstar}
\end{equation}
where $r_N$ is the nuclear length scale used to define the inner cutoff (Section~\ref{sec:refdata}). Around this scale the finite-nucleus multi-electron references flatten to a plateau, whereas the point-Coulomb hydrogen density has the weak $r^{2\gamma-2}$ singularity of equation~\eqref{eq:dirac1s}; neither limiting form can be represented by the finite polynomial-times-exponential basis. This structurally irreducible mismatch corresponds to $g$ values far beyond the fitted bandwidth. In the fitting implementation, the plain, $r^2$-weighted and log-density residuals start at the larger of this physical cutoff and a visibility threshold on $4\pi r^2\rho(r)$; the log--log slope and characteristic-radius terms use their own bulk and landmark masks.

\begin{table}[t]
\centering
\small
\caption{The components $\sigma_i$ of the cost function~\eqref{eq:cost} and their effective canonical weights $w_i$, grouped into the four active families used in the production fits. Each data and characteristic-radius term is a dimensionless normalised mean absolute deviation of the fitted curve from the relativistic reference; the physicality terms are penalties that vanish when the constraint holds. The listed weights include the global factor of 2000 used in the production configuration.}
\label{tab:cost}
\begin{tabular}{lc}
\toprule
Component $\sigma_i$ & $w_i$ \\
\midrule
\multicolumn{2}{l}{\emph{Data channels}}\\
$\quad$ $f_e$ residual                       & $2.67\times10^{3}$ \\
$\quad$ $f_e$, $g^2$-weighted (high-$g$ tail) & $6.36\times10^{2}$ \\
$\quad$ $\log f_e$ residual                  & $2.59\times10^{3}$ \\
$\quad$ $f_x$ residual                       & $6.40\times10^{2}$ \\
$\quad$ $f_x$, $g^2$-weighted                & $1.78\times10^{1}$ \\
$\quad$ $\log f_x$ residual                  & $2.05\times10^{1}$ \\
$\quad$ $\rho$ residual                      & $2.00\times10^{-2}$ \\
$\quad$ $\rho$, $r^2$-weighted ($4\pi r^2\rho$) & $4.80\times10^{-1}$ \\
$\quad$ $\log\rho$ residual                  & $1.27\times10^{-1}$ \\
$\quad$ log--log slope of $r^2\rho$          & $8.88\times10^{-2}$ \\
\midrule
\multicolumn{2}{l}{\emph{Physicality (zero when satisfied)}}\\
$\quad$ positivity of $f_e,f_x,\rho,V$       & $2.00\times10^{3}$ \\
$\quad$ monotone decay of $f_e,f_x,\rho,V$   & $2.00\times10^{3}$ \\
\midrule
\multicolumn{2}{l}{\emph{Characteristic radii of $4\pi r^2\rho$}}\\
$\quad$ peaks                                & $2.40\times10^{0}$ \\
$\quad$ shoulders                            & $2.40\times10^{0}$ \\
$\quad$ valleys                              & $2.40\times10^{0}$ \\
$\quad$ inflections                          & $4.00\times10^{-1}$ \\
$\quad$ rising-flank tail anchors            & $2.40\times10^{0}$ \\
$\quad$ falling-flank tail anchors           & $2.40\times10^{0}$ \\
\midrule
\multicolumn{2}{l}{\emph{Origin anti-cancellation regularisers}}\\
$\quad$ $\rho(0)$ balance                    & $1.84\times10^{-4}$ \\
$\quad$ $f_x(0)$ balance                     & $4.50\times10^{-6}$ \\
$\quad$ $f_e(0)$ balance                     & $5.65\times10^{-5}$ \\
\bottomrule
\end{tabular}
\end{table}

The weights were adjusted empirically so that no primary representation of the reference dominated the optimisation merely because of its numerical scale. Recomputed from the delivered Dirac--Pad\'e coefficients and averaged over all 118 elements, each of the ten data channels contributes between 7.5 and 8.6\% of the mean total cost; together they account for 76.7\%. The six characteristic-radius groups contribute 11.2\% and the three origin-balance regularisers 12.1\%, while all positivity and decay contributions are zero. These fractions describe the balance of the canonical objective rather than additional fitted observables.

\subsection{Optimisation and numerical conditioning}
\label{sec:optim}

The fitting parameters separate into the nonlinear widths $\{b_i\}$ and the linear amplitudes $\{a_i\}$. For a given set of widths, a constrained linear least-squares solve supplies the amplitudes on the exact-constraint manifold, and the full weighted cost~\eqref{eq:cost} is then evaluated for that coefficient pair. The outer optimisation searches over the widths. Relative to the earlier fitting procedure \cite{Lobato2014}, the numerical refinements described below improve robustness across all 118 elements.

\paragraph{Null-space inner solve} For fixed widths the amplitudes follow from the linearly constrained least-squares problem
\begin{equation}
\min_a \|F a - f\| \quad\text{subject to}\quad T a = c,
\label{eq:constrained_ls}
\end{equation}
where $F$ is the design matrix of the linear data channels, $f$ the stacked data-channel target vector, and $T a = c$ collects the active exact constraints (charge and $\langle r^2\rangle$ for hydrogen; charge, $\langle r^2\rangle$ and $\langle r^4\rangle$ otherwise). This least-squares solve is the inner linear surrogate used to place the amplitudes on the exact-constraint manifold; the outer optimiser still grades every trial by the full weighted cost~\eqref{eq:cost}. We solve the linear surrogate by the null-space method \cite{Press2007}: with $N$ an orthonormal basis for the null space of $T$ (the trailing columns of the orthogonal factor $Q$ in the QR factorisation of $T^{\mathrm{T}}$) and $a_p$ a particular solution of $Ta=c$, the amplitudes are
\begin{align}
a &= a_p + N\,y, \notag\\
y &= \arg\min_y \big\|F N\,y - (f - F a_p)\big\|.
\label{eq:nullspace}
\end{align}
The constraints then hold to floating-point precision by construction, and the forward error scales with the condition number of $T$ rather than with its square, as in the earlier block-elimination scheme, eliminating the near-singular inner solves that otherwise corrupt the cost landscape seen by the outer optimiser. The inner solve is carried out in double precision throughout---the two largest measured reduced-design condition numbers occur at Rg and Og (Section~\ref{sec:res_coef})---while the outer search quantises the width and amplitude parameters to single precision (float32) at every trial, so the delivered coefficients are single-precision by construction, matching the working precision of GPU multislice codes.

\paragraph{Ordered log-gap reparameterisation} Across the fitted periodic table, the widths span up to about $6.7$ decades in $b$, and two coinciding non-relativistic widths produce linearly dependent design columns whose amplitudes diverge with opposite signs---``ghost pairs'' that cancel in $f_e$ and $f_x$ but propagate catastrophic floating-point cancellation to any downstream consumer of $\{a_i\}$. We eliminate them by searching not over the non-relativistic widths directly but over a non-negative ordered log-gap slack vector $s$ (the search variable, with entries the log-gaps between successive widths), which maps the $n_{\mathrm{NR}}$ non-relativistic widths---within the per-atom box $[b_{\min}(Z),b_{\max}(Z)]$ of equation~\eqref{eq:bounds} below---by
\begin{equation}
\begin{aligned}
\log b_i &= \log b_{\max}-\sum_{j=1}^{i} s_j-(i-1)\,\delta_{\min},\\
i &= 1,\dots,n_{\mathrm{NR}}.
\end{aligned}
\label{eq:loggap}
\end{equation}
Here $n_{\mathrm{NR}}=n_t$ for the matched NR basis and $n_{\mathrm{NR}}=M=n_t-1$ for the Dirac--Pad\'e basis. The single Dirac--Pad\'e width belongs to a different functional family and is searched independently through
\begin{equation}
\log b_{\mathrm{DP}}=\log b_{\max}-s_{\mathrm{DP}},
\qquad 0\leq s_{\mathrm{DP}}\leq \log(b_{\max}/b_{\min}).
\label{eq:loggap_dp}
\end{equation}
Equation~\eqref{eq:loggap} is strictly decreasing and enforces $\log b_i - \log b_{i+1} \ge \delta_{\min}$ within the non-relativistic subset, so degenerate same-family configurations are unrepresentable. The Dirac--Pad\'e width remains free to cross the ordered non-relativistic widths because its design column has a different functional form. We use the empirically selected value $\delta_{\min} = 0.1$, which removed numerically degenerate non-relativistic width pairs without constraining the well-separated fitted widths. The per-atom width box $[b_{\min}(Z), b_{\max}(Z)]$ is built from the single-term identity
\begin{equation}
b = (\pi r_{\mathrm{peak}})^2
\label{eq:b_rpeak}
\end{equation}
that maps a hydrogenic width to the peak radius of its radial distribution. Applying the identity at the two extreme length scales of the atom gives
\begin{equation}
b_{\max} = B_{\mathrm{pad}}\,(\pi r_{\mathrm{outer}})^2, \qquad
b_{\min} = \frac{\gamma^2(Z)\,(\pi a_0)^2}{S\,Z^{p}},
\label{eq:bounds}
\end{equation}
where $r_{\mathrm{outer}}$ is the outermost characteristic radius of $4\pi r^2\rho(r)$. The lower bound is a search-box floor, not an assignment of any fitted term to a physical shell: it is obtained by applying the same width--radius identity to the relativistically contracted hydrogenic inner length scale $\gamma(Z)\,a_0/Z$ (for the hydrogenic exponent $p=2$). Because $\gamma^2(Z)=1-(Z\alpha)^2$ decreases as $Z$ increases, this factor lowers the floor and keeps the available width range open down to that explicit inner length scale. The padding factors are $B_{\mathrm{pad}}=10$ and $S=5$. The exponent is steepened empirically to $p \simeq 2.6$ to keep the lower bound below the inner-shell length scales measured from the conditioned relativistic reference densities described in Section~\ref{sec:refdata}; these choices define only the optimiser search box, not additional physical constraints on the fitted coefficients or on the location of the Dirac--Pad\'e term.

\begin{figure}[t]
\centering
\includegraphics[width=\textwidth]{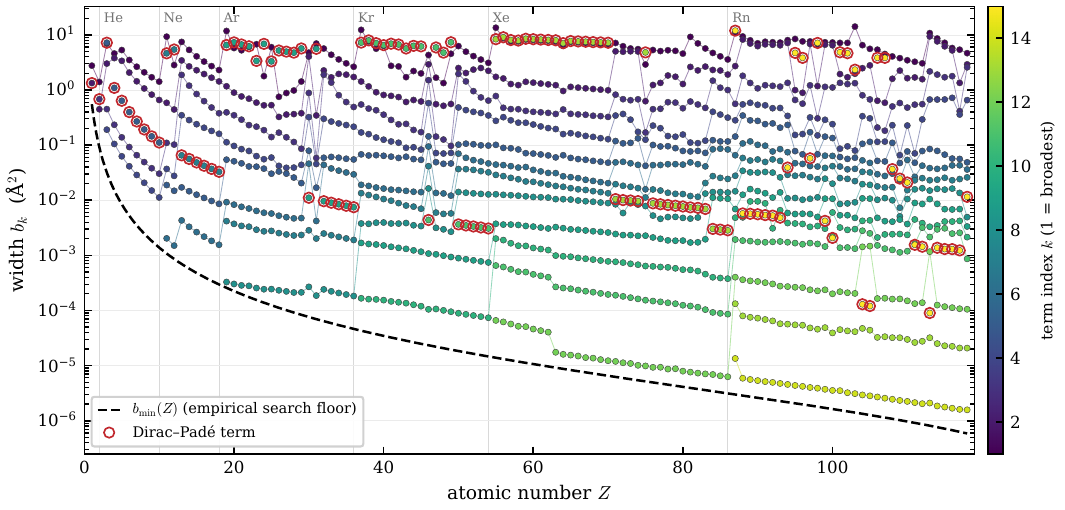}
\caption{Fitted nonlinear widths $b_k$ of the Dirac--Pad\'e basis for all 118
elements, coloured by term index ($k=1$ the broadest term). The widths span
about seven decades across the periodic table and grow in number at the
noble-gas boundaries (vertical lines, period symbols above), where the
element-adaptive basis size $n_t(Z)$ adds terms to resolve each newly opened
shell (Table~\ref{tab:nt}). The black dashed curve is the empirical lower search
bound $b_{\min}(Z)$ of equation~\eqref{eq:bounds}, below which no width is
placed; the fitted widths sit above this relativistically tightened floor. Red
rings mark the charge-carrying Dirac--Pad\'e term (the last basis term), whose width
is placed freely by the fit and is not part of the descending ordering; the
Dirac--Pad\'e label identifies the $n=3$ Pad\'e member rather than a physical shell
assignment.}
\label{fig:widths}
\end{figure}

Figure~\ref{fig:widths} collects the fitted widths for all 118 elements. The basis fans out at the noble-gas boundaries as $n_t(Z)$ grows, and every fitted width respects the $b_{\min}(Z)$ floor of equation~\eqref{eq:bounds}. The red-ringed Dirac--Pad\'e width is freely placed rather than tied to this lower bound; the label denotes the $n=3$ member selected by the Dirac--Pad\'e construction, not a component located at the physical inner-shell radius. The atoms are fitted independently, with no coupling or smoothness regularisation across $Z$. The plotted lines merely connect the element-by-element fitted widths as a visual guide and should not be interpreted as evidence that the constrained amplitude solve is well conditioned.

\paragraph{Global search} The fit is uncoupled across elements, so an individual atom can be re-optimised independently without disturbing its neighbours. The outer optimisation over the slack vector $s$ uses two complementary global optimisers, which fail on different atoms: simulated annealing in the Corana variant \cite{Corana1987} and a covariance-matrix-adaptation evolution strategy \cite{Hansen2001} run with a bi-population restart scheme (BIPOP-CMA-ES) \cite{Hansen2009}. In the annealing, each move proposes a one-coordinate trial, accepted with the Metropolis probability, and the temperature is lowered geometrically:
\begin{align}
s'_k &= s_k + (2u-1)\,v_k, \quad u\sim\mathcal{U}(0,1), \label{eq:sa_move}\\
P_{\mathrm{acc}} &= \min\!\left[1,\exp(-\Delta\chi^2/\mathcal{T})\right], \qquad
\mathcal{T} \leftarrow r_t\,\mathcal{T}. \label{eq:sa_metropolis}
\end{align}
The cooling runs from $\mathcal{T}_0$ to $\mathcal{T}_{\min}$, and each step size $v_k$ is adapted towards a target acceptance band as described by \citeasnoun{Corana1987}, with $\mathcal{T}_0 = 0.1$, $\mathcal{T}_{\min} = 10^{-7}$ and $r_t = 0.93$. At each temperature, 10 step-size updates are performed; each update comprises 20 Metropolis samples per coordinate, so one temperature level evaluates 200 samples per coordinate before cooling. BIPOP-CMA-ES starts with an isotropic standard deviation equal to 0.25 times the available slack range, uses nine restarts, and has a per-restart evaluation budget of $\max(10{,}000,1000n_s^2)$, where $n_s$ is the dimension of the slack vector. Both optimisers share the null-space inner solve~\eqref{eq:nullspace} and a terminal bound-constrained quasi-Newton (L-BFGS-B \cite{Byrd1995}) polish. Equation~\eqref{eq:loggap} orders only the non-relativistic widths, removing their equivalent permutations, while the Dirac--Pad\'e width follows the independent map~\eqref{eq:loggap_dp}. The production recipe begins with ten \emph{cold starts} per atom for each optimiser---independent random restarts from scratch---keeping the lower-cost result. Because the measured conditioning and cancellation vary strongly by element---with the largest reduced-design condition numbers at Rg and Og rather than at a universal atomic-number boundary (Section~\ref{sec:res_coef})---this initial pass is refined by three kinds of targeted pass: \emph{warm starts}, which re-optimise an atom from its current best solution; \emph{coefficient-} and \emph{cost-outlier rescues}, which refit the atoms whose fitted widths $\{b_i\}$ depart from the smooth trend with $Z$ or whose cost $\chi^2$ stands above those of their neighbours with the same $n_t$; and \emph{neighbour-seeded explorations}, which sweep up and then down the periodic table, reseeding each atom from its already-fitted neighbour. Every pass accepts a candidate only when it strictly lowers $\chi^2$, so the stored fit improves monotonically.

\paragraph{Delivered-precision physicality acceptance} The soft penalties in equation~\eqref{eq:realspace_soft_positivity} shape the cost landscape but do not establish admissibility. Before selection or storage, the coefficients are cast to the declared delivery precision $p$ ($p=32$ for the production tables), the maximum relative residual of the active exact constraints must not exceed $10^{-6}$, and $\rho$ and $W=rV$ must pass a deterministic hard test. This test searches $[0,r_{\max}]$, including adaptively resolved stationary points, and separately requires a positive analytic tail after grouping exactly coincident delivered widths. Its margin accounts for summation round-off at precision $p$ and the discrepancy from float64 evaluation of the delivered coefficients. An intermediate negative interval, negative analytic tail or delivery-precision sign loss therefore causes rejection. The hard gate applies only to these two real-space sign constraints; reciprocal-space positivity and all four monotone-decay conditions remain sampled objective terms.

Post-fit validation of the released coefficients uses exact polynomial root isolation in reciprocal space and overlapping component and asymptotic bounds in real space. The validation script and machine-readable results accompany the released data.

\section{Reference data}
\label{sec:refdata}

The reference electron densities for the multi-electron elements $Z=2$--$118$ are all-electron relativistic Dirac--Fock densities computed with the \texttt{atomx} code \cite{Zhang2026atomx}. The hydrogen reference is constructed from the exact point-Coulomb Dirac $1s$ density of equation~\eqref{eq:dirac1s}, stored in the same radial-data convention, and passed through the common conditioning and transform pipeline. The same exact solution provides the analytical model for the Dirac--Pad\'e construction in Section~\ref{sec:basis}. The multi-electron calculations use the tabulated neutral ground configuration and a spherical $jj$-coupled open-shell average. For configurations containing multiple $jj$-coupled configuration-state functions, \texttt{atomx} obtains the common radial orbitals from its Hamiltonian-weighted average and the density records use the statistical $(2j+1)$-degeneracy split for the final subshell occupations; closed shells reduce to a direct Dirac--Fock solve. The complete \texttt{atomx} package is maintained separately by Zhang and is not redistributed with the present parameterisation.

For the multi-electron \texttt{atomx} calculations, the radial equations are represented by 113 B-splines of order 9 on a DBSR\_HF-style semi-exponential grid. For the finite Fermi nucleus this grid starts at $1.09\times10^{-5}a_0$, grows by a factor of 1.25 until the interval reaches $a_0$, and then continues with constant $a_0$ spacing to a cavity radius of $50.54a_0$. The self-consistent-field iteration uses a maximum of 300 iterations and a relative orbital tolerance of $10^{-6}$, together with an energy-change threshold of $10^{-8}$; a Thomas--Fermi--Dirac--Amaldi starting potential is used for $Z\geq20$. The nuclear Fermi thickness is \SI{2.30}{\femto\metre}. For $2\leq Z\leq10$, tabulated root-mean-square charge radii are used subject to the DBSR\_HF lower bound of \SI{2.0}{\femto\metre}; consequently, the tabulated helium value of \SI{1.6755}{\femto\metre} is replaced by \SI{2.0}{\femto\metre}. Above this range, the DBSR\_HF empirical radius is used with mass number $A=\operatorname{round}(2.5Z)$. Breit, vacuum-polarisation and self-energy corrections are evaluated perturbatively after convergence. They contribute to the reported atomic energies but are not fed back into the radial orbitals and therefore do not alter the reference density used in this work.

The total electron density follows from the large ($P_a$) and small ($Q_a$) radial components of the occupied orbitals,
\begin{equation}
4\pi r^2\rho(r) = \sum_a n_a\left[P_a(r)^2 + Q_a(r)^2\right],
\label{eq:rho_components}
\end{equation}
which integrates to $Z$, so the relativistic contraction of the inner shells enters directly. Each subshell is stored on the native Gauss--Legendre quadrature grid and trimmed after its large-component contribution falls below $10^{-17}$; the converter sums these exact grid prefixes with zero padding and converts the radial-distribution convention of \texttt{atomx} to $\rho(r)$. The native Gauss--Legendre nodes lie inside their integration intervals and therefore do not include the endpoint $r=0$; the smallest tabulated radius is \SI{9.15e-8}{\angstrom}. The first two volume-density samples are conditioned with a log-quadratic curve fitted to the next three samples and anchored continuously at the first retained value; the same curve supplies the explicit $r=0$ sample required by the transform and moment grids. The assigned finite origin value does not change the transform or radial moments because their radial integrands contain $4\pi r^2\rho(r)$, which vanishes at $r=0$. The continuation inserts the origin and replaces only the first two raw samples; no original sample beyond $r=\SI{0.04714}{\femto\metre}$ is modified. This conditioning does not impose a cusp condition on the fitted parameterisation.

The large-radius density should become log-linear once the outermost orbital dominates. The finite B-spline calculation instead reaches a numerical floor for some elements, after which the tabulated tail flattens or oscillates. A local-slope test identifies the onset of this floor, a log-linear model is fitted to the preceding clean asymptotic interval, and only the affected low-density tail is continued to the common cutoff $\rho=\SI{1e-30}{\per\angstrom\cubed}$. Finally, the conditioned density is rescaled uniformly so that $\int 4\pi r^2\rho(r)\,dr=Z$ on the $2^{20}$-point quadrature grid. The conditioned coarse mesh is then retained without further resampling as the real-space fitting target. These conditioning steps are applied to every element, including hydrogen. The reciprocal-space and moment quantities used below are computed in this work from this conditioned density.

The reciprocal-space targets $f_e(g)$ and $f_x(g)$ are computed numerically from the reference density rather than taken from an external tabulation. The density is first cubic-spline interpolated in $\log\rho$ onto a $2^{20}$-point logarithmic radial grid; $f_x(g)$ is then obtained from the Hankel transform of equation~\eqref{eq:fxg} on that grid and $f_e(g)$ from the Mott--Bethe formula~\eqref{eq:feg}, with the $g=0$ values fixed to $f_x(0)=Z$ and to the Ibers limit~\eqref{eq:fe0}. The fine sampling is essential rather than cosmetic: the transform kernel $\sin(2\pi g r)/(2\pi g r)$ oscillates ever faster as $g$ grows, so on a coarse mesh the quadrature aliases and the computed $f_x(g)$ rings; the $2^{20}$-point grid resolves the kernel out to the full \SI{36}{\per\angstrom} without spurious oscillation. The $g$-grid is extended to \SI{36}{\per\angstrom} (721 points at a \SI{0.05}{\per\angstrom} step), well beyond the \SI{12}{\per\angstrom} range of the tabulated scattering factors \cite{Kirkland1998}, which was also the upper limit of the earlier fit \cite{Lobato2014}, to cover aberration-corrected instruments, higher-order Laue zones (HOLZ) in quantitative CBED, and thermal diffuse scattering backgrounds in frozen-phonon multislice \cite{LobatoMULTEM2015,VanDyck2009}. The moments $\langle r^n\rangle$ and $p_n = Z\langle r^n\rangle$ are computed from the reference density by quadrature on the same fine grid; the $n=0$ case recovers the atomic number $Z$ to numerical roundoff, and the $n=2$ case fixes the Ibers value $f_e(0)=p_2/(3a_0)$ used as the $g=0$ anchor of the Mott--Bethe inversion. Independent relativistic Dirac--Hartree--Fock tabulations \cite{Olukayode2023} and the Dirac--Fock densities reported by \citeasnoun{Guerra2017} were used as qualitative cross-checks on the transform and moment pipeline rather than as additional fitting targets.

\section{Results}
\label{sec:results}

We fit all 118 neutral atoms independently with both the Dirac--Pad\'e and the matched NR basis under identical weights, so that the two differ only in the basis functional form at fixed parameter count; hydrogen, now fit to the relativistic reference like every other element, takes two terms, and the matched parameter count $2n_t$ ranges from $4$ (H) to $30$ (period 7) following the schedule of Section~\ref{sec:adaptive}. The Dirac--Pad\'e-versus-NR comparison below covers all 118 elements ($Z = 1$--$118$). The fixed five-term refit, labelled `Fixed-5', serves as the controlled baseline for isolating the effect of the element-adaptive basis. It uses the five-term basis size of the published parameterisation \cite{Lobato2014} but extends its 103-element scope to all 118 elements under the present reference, grid, constraints and cost; its original coefficients are neither reused nor rescored.

\subsection[Accuracy of fe and fx]{Accuracy of $f_e(g)$ and $f_x(g)$}
\label{sec:res_feg}

\begin{figure}[t]
\centering
\includegraphics[width=\textwidth]{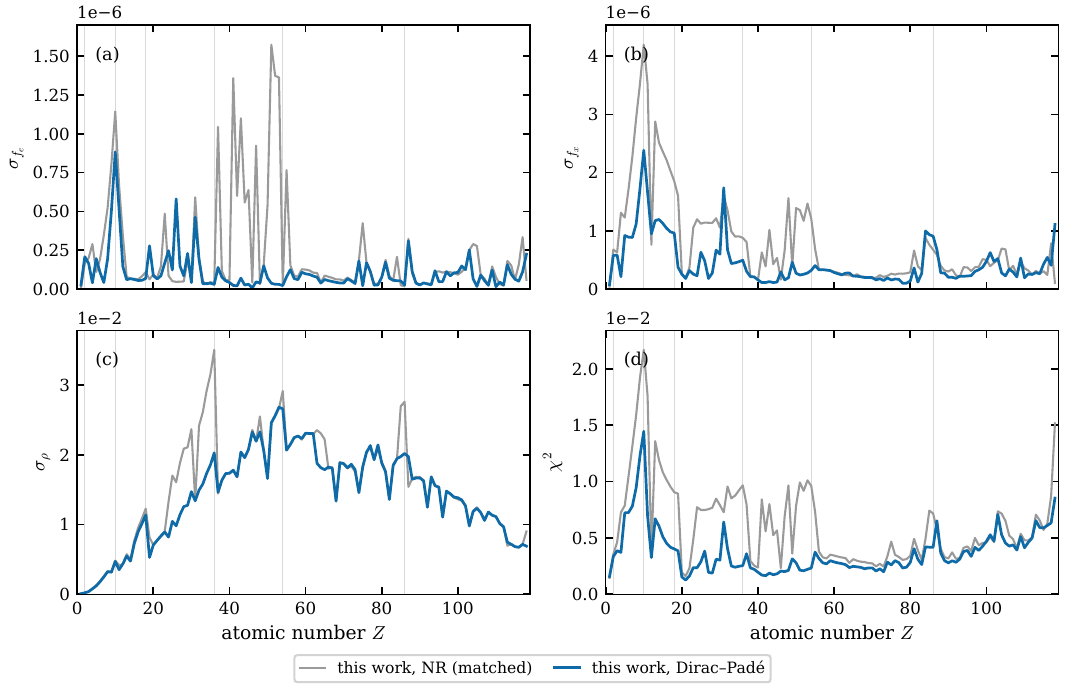}
\caption{The normalised mean absolute deviations $\sigma_{f_e}$, $\sigma_{f_x}$ and $\sigma_{\rho}$, together with the weighted total cost $\chi^2$, are plotted against atomic number $Z$ for the Dirac--Pad\'e parameterisation of this work and the parameter-matched non-relativistic baseline. The first three panels are post-fit diagnostics, while $\chi^2$ is the scalar objective minimised during fitting. The fixed five-term refit is omitted from the plot so that the two adaptive bases can be compared on their natural scale; its much larger deviations are given in Table~\ref{tab:headline_accuracy}. The density deviation $\sigma_{\rho}$, measured outside the masked point-nucleus region and normalised by the peak of the represented density, sits well above the reciprocal-space deviations because the steep near-nuclear density is governed by the weakly constrained high-$g$ tail of the scattering factors. Vertical lines mark the noble-gas period boundaries.}
\label{fig:sigma}
\end{figure}

Figure~\ref{fig:sigma} reports the normalised mean absolute deviations $\sigma_{f_e}$, $\sigma_{f_x}$ and $\sigma_{\rho}$ for every element, together with the total weighted cost $\chi^2$, for the two parameter-matched adaptive bases. Table~\ref{tab:headline_accuracy} lists the corresponding medians and ranges, and retains the fixed five-term refit as the controlled fixed-size baseline. The three $\sigma$ channels are post-fit diagnostics, not independent minimisation targets: each coefficient record is selected by minimising the total weighted cost $\chi^2$ of equation~\eqref{eq:cost}. The Dirac--Pad\'e parameterisation reproduces the electron scattering factor to a median $\sigma_{f_e}=\num{6.63e-8}$ and the X-ray scattering factor to a median $\sigma_{f_x}=\num{2.79e-7}$ across the full periodic table. A direct numerical comparison with the published coefficients \cite{Lobato2014} is not used for the headline metrics, since those coefficients were fitted to a different reciprocal-space target and range. The controlled comparison is the Fixed-5 refit: the same five-term basis size, extended from the published 103-element scope to all 118 elements and refitted to the present relativistic reference under the present cost. Holding the basis at this fixed five-term size gives medians $\sigma_{f_e}=\num{8.33e-5}$ and $\sigma_{f_x}=\num{1.36e-3}$; the element-adaptive Dirac--Pad\'e basis, fitted under the same cost against the same reference, instead reaches \num{6.63e-8} and \num{2.79e-7}. The fixed five-term basis is therefore the binding constraint: five terms cannot resolve the progressively increasing shell structure as additional principal shells are occupied, which is exactly the limitation the element-adaptive basis size (Section~\ref{sec:adaptive}) is designed to remove.

The density panel reports $\sigma_{\rho}$, formed on the same footing but in real space and over the physically representable region outside the masked point-nucleus cutoff $r^*$ (Section~\ref{sec:cost}), normalised by the peak of the represented density. Both present bases reproduce the density to a median $\sigma_{\rho}=\num{1.49e-2}$ for the Dirac--Pad\'e basis and $\num{1.65e-2}$ for the matched NR basis, modestly tighter than the fixed five-term refit ($\sigma_{\rho}=\num{2.96e-2}$ over all 118 elements); the decisive density advantage of the element-adaptive construction over Fixed-5 is the shell-resolved structure (Figures~\ref{fig:targets_light} and~\ref{fig:targets_heavy}), not this element-averaged metric. The density deviation sits well above $\sigma_{f_e}$ and $\sigma_{f_x}$ because the quantities are connected by an ill-conditioned inverse problem: the density at small $r$ is set by the scattering factor at large $g$ through the Hankel transform, and that high-$g$ tail is the least-constrained part of any finite parameterisation. The Dirac--Pad\'e and matched non-relativistic bases are comparably accurate in the three diagnostic channels of Figure~\ref{fig:sigma}, with the Dirac--Pad\'e basis modestly tighter on $f_x$; their difference is expressed most clearly by the total weighted cost analysed in Section~\ref{sec:res_dirac_pade}.

\begin{table}[t]
\centering
\scriptsize
\setlength{\tabcolsep}{1pt}
\caption{Headline deviations evaluated from the stored single-precision coefficients. The first three rows are unweighted post-fit diagnostics; the final row is the weighted total cost $\chi^2$ of equation~\eqref{eq:cost}, which is the minimised objective. Dirac--Pad\'e and matched NR cover all 118 elements; Fixed-5 is the five-term refit to the same reference, now extended to all 118 elements. Each entry gives median (minimum--maximum).}
\label{tab:headline_accuracy}
\begin{tabular}{lccc}
\toprule
Metric & Dirac--Pad\'e & matched NR & Fixed-5 \\
\midrule
$\sigma_{f_e}$ &
\num{6.63e-8} (\num{7.52e-9}--\num{8.84e-7}) &
\num{1.03e-7} (\num{6.96e-9}--\num{1.57e-6}) &
\num{8.33e-5} (\num{8.55e-9}--\num{6.29e-4}) \\
$\sigma_{f_x}$ &
\num{2.79e-7} (\num{7.62e-8}--\num{2.38e-6}) &
\num{3.99e-7} (\num{4.22e-8}--\num{4.20e-6}) &
\num{1.36e-3} (\num{1.97e-8}--\num{2.04e-3}) \\
$\sigma_{\rho}$ &
\num{1.49e-2} (\num{7.78e-5}--\num{2.69e-2}) &
\num{1.65e-2} (\num{8.43e-5}--\num{3.51e-2}) &
\num{2.96e-2} (\num{7.73e-5}--\num{1.03e-1}) \\
$\chi^2$ &
\num{3.06e-3} (\num{1.26e-3}--\num{1.45e-2}) &
\num{4.96e-3} (\num{1.45e-3}--\num{2.17e-2}) &
\num{3.87e0} (\num{1.18e-3}--\num{6.87e0}) \\
\bottomrule
\end{tabular}
\end{table}

\clearpage

\subsection{Dirac--Pad\'e versus matched non-relativistic basis}
\label{sec:res_dirac_pade}

The global diagnostics leave one basis-level question unresolved. The Fixed-5 refit in Table~\ref{tab:headline_accuracy} shows that a fixed five-term basis is no longer sufficient, but it does not by itself isolate the value of the Dirac--Pad\'e term because Fixed-5 also has fewer parameters. The relevant comparison is therefore the pair already plotted in Figure~\ref{fig:sigma}: two matched models under the same weights, constraints and $n_t(Z)$ schedule, namely the Dirac--Pad\'e basis with $M=n_t-1$ non-relativistic terms plus one Dirac--Pad\'e term, and the matched NR basis with $n_t$ non-relativistic terms.

The lower-right panel of Figure~\ref{fig:sigma} gives the element-by-element visual result, and Table~\ref{tab:headline_accuracy} gives the corresponding median and range of the minimised total cost $\chi^2$. We use the \SI{2}{\percent} band only as a classification tolerance: ratios inside it are treated as ties, not as a bound on the improvement. By this criterion, the Dirac--Pad\'e basis has a clearly lower total weighted cost for 111 elements; one element is a tie, and the matched NR basis is lower for six. The largest matched-NR advantage is \SI{14}{\percent} at Fr after imposing delivered-precision physicality; the other five matched-NR wins differ by at most about \SI{9}{\percent}. The mean total cost falls from \num{6.15e-3} (matched NR) to \num{3.75e-3} (Dirac--Pad\'e), a \SI{39}{\percent} reduction. The median per-element ratio $\chi^2_{\mathrm{NR}}/\chi^2_{\mathrm{DP}}$ is 1.23, so the matched-NR cost is typically \SI{23}{\percent} higher, or equivalently the Dirac--Pad\'e cost is about \SI{19}{\percent} lower relative to matched NR. The ratio of the two median costs in Table~\ref{tab:headline_accuracy} is 1.62 because the medians are formed separately. The upper tail is not unique to Nb: the five largest observed ratios are 4.95 for Nb, 4.66 for Sb, 4.58 for I, 4.56 for Cd and 4.40 for Te. At Nb, the improvement is distributed across several real- and reciprocal-space channels of the common objective. The present data therefore identify an empirical maximum but do not establish a unique Nb-specific physical mechanism.

The fitted Dirac--Pad\'e coefficients also confirm that the term is not a physical shell decomposition. The amplitude $a_{\mathrm{DP}}$ is negative for 59 elements and positive for 59; its sign carries no separate physical charge, because all amplitudes are signed expansion coefficients (Section~\ref{sec:dirac_pade}). The fitted widths give the same message: for $Z\ge2$ the Dirac--Pad\'e width is never the narrowest width, but is the broadest for 39 atoms and lies between non-relativistic widths for the remaining 78. The Dirac--Pad\'e term is therefore a freely placed shape channel, not a fitted component at the physical inner-shell radius.

\subsection{Shell-resolved electron density}
\label{sec:res_rho}

\begin{figure}[tp]
\centering
\includegraphics[width=\textwidth]{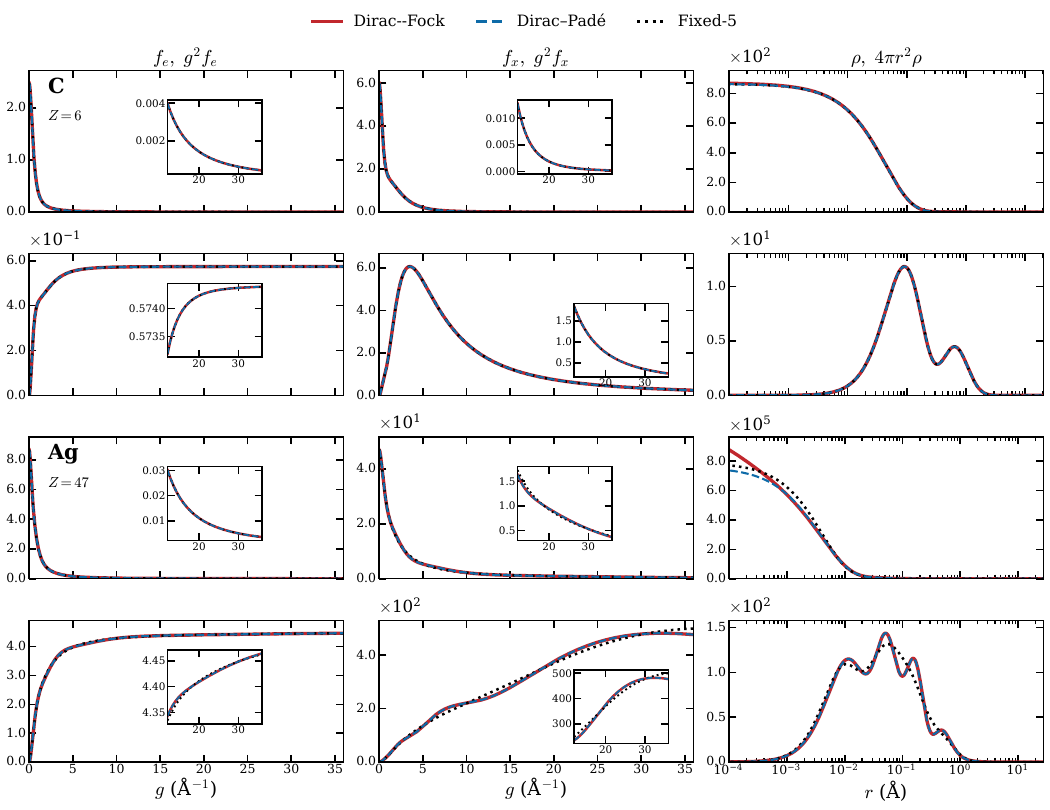}
\caption{The quantities entering the fit, shown for C and Ag, each as a $2\times 3$ block: the electron scattering factor $f_e(g)$ and its $g^2$-weighted form (left column), the X-ray scattering factor $f_x(g)$ and its $g^2$-weighted form (centre), and the electron density $\rho(r)$ and its radial distribution $4\pi r^2\rho(r)$, both on a logarithmic radial axis (right). Each reciprocal panel carries a high-$g$ inset ($g\ge\SI{12}{\per\angstrom}$). Within each element's block the upper row is the plain quantity and the lower row its $g^2$-weighted form (reciprocal columns) or the radial distribution $4\pi r^2\rho(r)$ (density column). The Dirac--Fock reference is solid red, the Dirac--Pad\'e parameterisation is dashed blue, and the fixed five-term refit (`Fixed-5') is dotted black; the redundant line styles preserve the comparison in greyscale.}
\label{fig:targets_light}
\end{figure}

Beyond the total cost, the present real-space terms expose the shell structure to every model in the comparison, including Fixed-5. The fixed five-term basis still cannot follow that structure once the number of occupied shells grows, so its failure is a capacity limit under the present objective, not a recurrence of the earlier reciprocal-space-only fit. Figures~\ref{fig:targets_light} and~\ref{fig:targets_heavy} use C, Ag, Au and U as illustrative examples spanning periods 2, 5, 6 and 7 and adaptive basis sizes 5, 11, 13 and 15, respectively; they were not selected by a statistical ranking. The comparison shows that the Dirac--Pad\'e parameterisation tracks the visible radial shell peaks, whereas Fixed-5 collapses to a smoother envelope that misses inner-shell structure for Au and U. The residual is largest in the masked point-nucleus region, where the finite-nucleus reference flattens but the point-nucleus exponential basis cannot; the bare value $\rho(0)$ can differ substantially from the finite-nucleus reference across the later periods. This structurally irreducible mismatch is confined to radii below the fitting mask (Section~\ref{sec:cost}) and leaves the shell structure shown in Figures~\ref{fig:targets_light} and~\ref{fig:targets_heavy} unaffected.

\subsection[High-g tail and near-nuclear density]{High-$g$ tail and near-nuclear density}
\label{sec:res_allspaces}

\begin{figure}[tp]
\centering
\includegraphics[width=\textwidth]{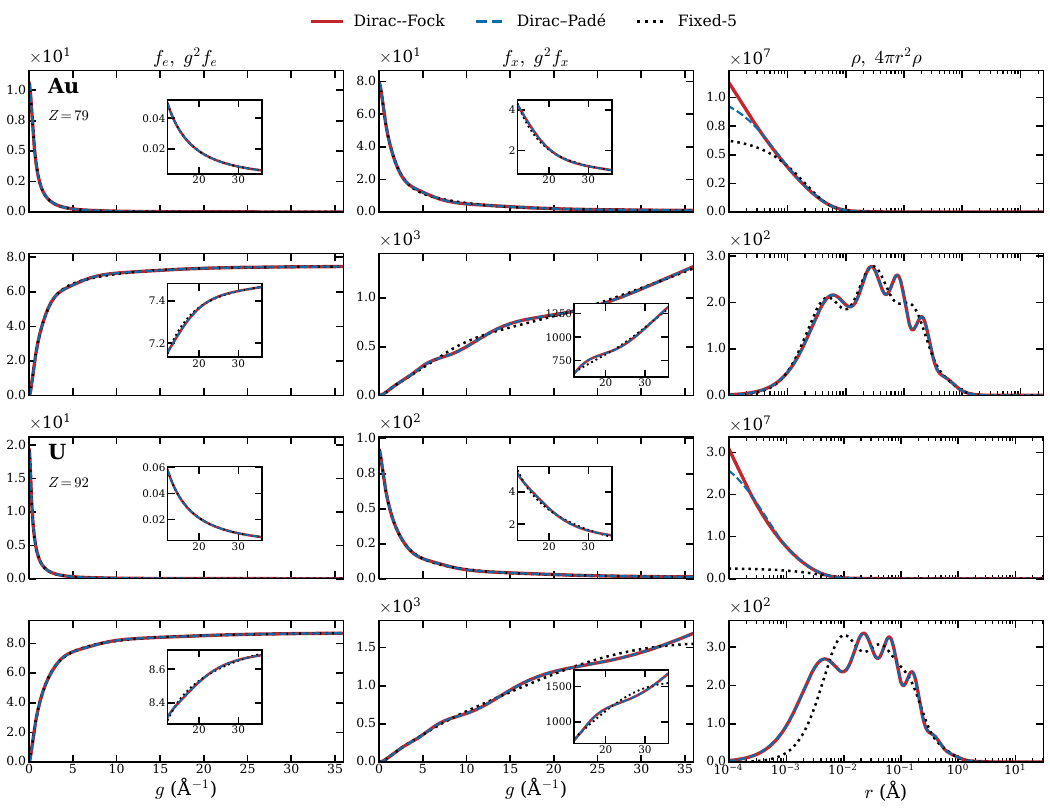}
\caption{The quantities entering the fit, shown for Au and U, with the same $2\times 3$ block layout, columns, curve identities and high-$g$ insets ($g\ge\SI{12}{\per\angstrom}$) as Figure~\ref{fig:targets_light}. The fixed five-term refit departs at the inner K, L, M and N shells of $4\pi r^2\rho(r)$ and develops percent-level oscillatory residuals in the high-$g$ tail of $f_x(g)$.}
\label{fig:targets_heavy}
\end{figure}

The high-$g$ tail links the reciprocal-space fit to the near-nuclear electrostatic potential. The Mott--Bethe nuclear term makes the relative deviation in $f_e(g)$ a weak diagnostic of near-nuclear-density errors, whereas $f_x(g)$ probes the electron density directly and exposes the inner-shell structure. Figure~\ref{fig:targets_heavy} therefore shows the $g^2$-weighted scattering factors next to the radial distribution: in the high-$g$ part of $f_x(g)$ the fixed five-term refit develops percent-level oscillatory residuals---the reciprocal-space counterpart of the missed inner shells of Section~\ref{sec:res_rho}---while the Dirac--Pad\'e parameterisation remains on the Dirac--Fock reference in both spaces. Tabulations limited to \SI{12}{\per\angstrom}---the range to which the earlier fit was restricted, and of most legacy data---leave this tail only indirectly constrained. The present high-$g$ accuracy follows from fitting the reference data directly to \SI{36}{\per\angstrom} while retaining the correct analytic asymptotics. The near-nuclear electrostatic potential $V(r)$ is built from the same coefficients through equation~\eqref{eq:work_Vr}, so errors in the high-$g$ tail propagate to $V(r)$ at small radii.

\subsection{Constraint satisfaction and physical validity}
\label{sec:res_cstr}

Recomputing the active exact constraints from the delivered single-precision coefficients gives maximum relative residuals of $2.55\times10^{-7}$ for Dirac--Pad\'e, $4.48\times10^{-7}$ for matched NR and $4.97\times10^{-7}$ for Fixed-5, all below the $10^{-6}$ acceptance limit. Canonical re-scoring gives zero sampled positivity and monotone-decay penalties on $f_e$, $f_x$, $\rho$ and $V$ for every record in all three sets, and all 118 selected Dirac--Pad\'e records pass the hard real-space test of Section~\ref{sec:optim}. The ordered log-gap reparameterisation also leaves no degenerate width pairs.

For the released float32 coefficients, these tests establish positivity of $f_e$, $f_x$, $\rho$ and $V$ throughout their physical domains for every element in all three fitted models. Positivity of the projected potential $V(R)$ follows from its line integral through $V(r)$. The finite real-space enclosures overlap the analytic tail bounds, so this result does not rely on sampled grids alone.

\subsection{Parameterised coefficients in single precision}
\label{sec:res_coef}

The fitted coefficients $\{a_i, b_i\}$ for all 118 elements are listed in full in Table~\ref{tab:coef_full} and provided as a machine-readable file in the data deposit, in single precision (float32), the working precision of GPU multislice codes. For the Dirac--Pad\'e basis the final term of each element (marked DP) is the charge-carrying Dirac--Pad\'e term and the remainder are non-relativistic (the amplitudes are signed expansion coefficients, Section~\ref{sec:dirac_pade}).

The exact constraints and the wide range of fitted length scales render the linear amplitude system ill-conditioned. The X-ray scattering factor must satisfy two endpoint conditions, $f_x(0)=Z$ and $f_x(g)\to 0$ as $g\to\infty$, and reproducing this vanishing high-$g$ tail with basis widths that span several decades forces large opposite-sign amplitudes that nearly cancel. At fixed widths the amplitudes enter $f_e(g)$ and $f_x(g)$ linearly. We quantify the constrained linear problem by $\kappa_2(\mathbf{A}\mathbf{N})$, where $\mathbf{A}$ is the active weighted linear design matrix and the columns of $\mathbf{N}$ span the null space of the exact-constraint matrix $\mathbf{T}$ in equation~\eqref{eq:nullspace}. In the Dirac--Pad\'e production run this reduced-design condition number reaches \num{9.07e8} at Rg and \num{4.28e8} at Og; independently, the charge-cancellation ratio $\sum_i|a_i/b_i|/|\sum_i a_i/b_i|$ reaches 124 at Og. These diagnostics explain why coefficients rounded only after a double-precision fit can lose substantial accuracy, but their element rankings need not match the observed float32 cost penalty.

Fitting directly in the precision of deployment avoids this loss: every trial is quantised to float32 with only the constrained inner solve in double precision (Section~\ref{sec:optim}), so the delivered coefficients are single-precision by construction. Re-scoring paired delivered float32 and float64 records changes the total fit cost by a median of only $0.065\%$ for all 118 Dirac--Pad\'e records and $0.081\%$ for all 118 matched-NR records. The penalty exceeds \SI{1}{\percent} for Dirac--Pad\'e only at Cn (1.35\%), Nh (2.62\%), Fl (1.34\%), Lv (1.60\%) and Og (1.94\%); for matched NR it exceeds \SI{1}{\percent} at Ta (1.14\%), Bk (1.12\%), Md (1.56\%), Db (1.05\%), Mt (1.27\%), Rg (2.30\%), Cn (2.84\%), Fl (1.19\%) and Mc (1.01\%). The effect is therefore concentrated mainly, but not exclusively, in later-period atoms and is numerical rather than evidence of a new element-specific physical effect.

The same cancellation requires care when the analytic terms are evaluated in single precision. A direct left-to-right sum loses low-order bits whenever a large positive contribution is followed by a nearly equal negative contribution. Compensated summation, such as the Kahan algorithm \cite{Kahan1965}, maintains a running correction for the rounding discarded at each addition and substantially reduces this arithmetic loss at negligible cost for at most 15 terms. It cannot recover information already removed by quantising the coefficients, which is why the fit itself is performed with float32-quantised trials; it protects only the subsequent evaluation of the signed term sum. Single-precision implementations should therefore use compensated accumulation, or accumulate the terms in double precision when that is available.

\section{Discussion}
\label{sec:discussion}

The element-adaptive parameterisation addresses the three principal limitations of the Lobato--Van Dyck (2014) approach while preserving its central advantage: all derived quantities---$f_x(g)$, $\rho(r)$, $V(r)$, and $V(R)$---remain available in closed analytic form. Simulation codes that currently accept that parameterisation require only two structural changes: reading $n_t(Z)$ from the parameter table and evaluating the additional Dirac--Pad\'e term.

\paragraph{Role of the Debye--Waller factor} At room temperature, an isotropic Debye--Waller amplitude factor $\exp(-B g^2/4)$ with a representative $B=\SI{0.5}{\angstrom\squared}$ attenuates coherent Bragg amplitudes by about 86\% at $g = \SI{4}{\per\angstrom}$ and by more than 99.9\% at $g = \SI{8}{\per\angstrom}$, which might seem to remove any practical value in improving scattering factors at these large reciprocal-space magnitudes. The objection applies only to the coherent channel. Thermal vibration splits the scattered wave into a time-averaged elastic (Bragg) component and a fluctuating diffuse component, and the recorded intensity is their incoherent sum \cite{VanDyck2009}. The Debye--Waller factor is the Fourier transform of the atomic-displacement distribution and multiplies only the time-averaged potential, so it damps the elastic component alone; the intensity it removes is not lost but conserved, equal to the total thermal diffuse scattering (TDS) generated by the fluctuating potential \cite{VanDyck2009}. This TDS is described either by an imaginary absorptive (optical) potential that represents the absorption of the elastic wave \cite{WeickenmeierKohl1991,Thomas2024}, or by the frozen-phonon average of displaced-atom snapshots, from which the Debye--Waller damping instead emerges \cite{LobatoMULTEM2015,VanDyck2009}---both resting on the same independent-atom (Einstein) thermal average and both built from the undamped high-$g$ scattering factor. The diffuse channel---which dominates the HAADF-STEM signal \cite{Thomas2024}---is moreover governed by a sharply peaked scatterer: the fluctuating potential is strongly peaked at the nucleus, so the vibrating atom acts as a near-point scatterer for the diffuse intensity, whose Fourier transform spans a large fraction of reciprocal space; by the Mott--Bethe large-$g$ asymptotic ($\sim\!Z/g^2$) the diffuse intensity integrated beyond a cutoff $g_{\max}$ then falls only as $Z^2/g_{\max}^2$ \cite{VanDyck2009}. Accurate high-$g$ scattering factors therefore enter the TDS background through the undamped electrostatic potential, rather than being suppressed by the coherent Debye--Waller factor in the same way as Bragg amplitudes.

\section{Conclusions}
\label{sec:conclusions}

We have computed updated all-electron relativistic Dirac--Fock reference densities with \texttt{atomx} for the multi-electron elements $Z=2$--$118$, supplemented them with the exact relativistic Dirac $1s$ hydrogen reference, and presented an element-adaptive parameterisation fitted to the complete 118-element set---of elastic electron and X-ray scattering factors, electron densities, and electrostatic potentials---retaining the closed-form analytic framework of the earlier hydrogenic parameterisation \cite{Lobato2014}. The extension adds four ingredients: an element-adaptive basis size, simultaneous real-space and reciprocal-space fitting against modern relativistic Dirac reference densities, an exact $\langle r^4\rangle$-moment constraint in place of the Kato cusp, and a charge-carrying Dirac--Pad\'e term that adds a polynomial-times-exponential shape channel while preserving closed-form derived quantities. The parameterisation reproduces $f_e$ and $f_x$ about three to four orders of magnitude more accurately than a five-term refit to the same reference and objective across all 118 elements, resolves shell structure in $4\pi r^2\rho(r)$ that five terms cannot, and improves on the parameter-matched non-relativistic basis for 111 of the 118 elements, lowering the mean total cost by \SI{39}{\percent}.

A separate mathematical analysis of the released float32 coefficients establishes that $f_e(g)$, $f_x(g)$, $\rho(r)$ and $V(r)$ remain positive throughout their physical domains for all 118 elements.

Several follow-ups are opened directly by this work. The updated analytic $V(r)$ can be tested as a replacement for the three-Yukawa approximation used in partial-wave Mott cross-section codes \cite{Salvat2005,Salvat2021}, with possible benefit to BSE imaging and EBSD pattern simulation \cite{Callahan2013,Winkelmann2021}. The more accurate IAM baseline is also useful for bonding-sensitive refinements---kappa refinement, Hirshfeld atom refinement, and the transferable aspherical atom model \cite{KappaRefinement2024,Chodkiewicz2024,Gruza2024}---against three-dimensional electron diffraction \cite{Mahmoudi2025} and 4D-STEM electron ptychography \cite{Hofer2025}. Machine-readable float32 coefficient tables for all three models accompany this paper (see Data and code availability); the same closed-form expressions also serve Bloch-wave simulation and any other method built on analytic neutral-atom scattering factors.

\appendix
\section{Full parameterisation table}
\label{app:coef}

\begingroup
\scriptsize
\setlength{\tabcolsep}{1.2pt}
\setlength{\LTpre}{0pt}
\setlength{\LTpost}{0pt}
\begin{longtable}{llcrrrrrr}
\caption{Fitted Dirac--Pad\'e coefficients for all 118 elements, in single precision (float32). For each element the $(a_i, b_i)$ pairs run left to right then top to bottom; the non-DP pairs are in order of decreasing width $b_i$, and the final pair is the charge-carrying Dirac--Pad\'e term, whose width is independently optimised and is not part of that ordering. $a_i$ in \si{\angstrom}, $b_i$ in \si{\angstrom\squared}.}\label{tab:coef_full}\\
\toprule
$Z$ & El. & $n_t$ & $a_i$ & $b_i$ & $a_i$ & $b_i$ & $a_i$ & $b_i$ \\
\midrule
\endfirsthead
\toprule
$Z$ & El. & $n_t$ & $a_i$ & $b_i$ & $a_i$ & $b_i$ & $a_i$ & $b_i$ \\
\midrule
\endhead
\midrule \multicolumn{9}{r}{continued on next page}\\
\endfoot
\bottomrule
\endlastfoot
1 & H & 2 & $2.6462868 \times 10^{-1}$ & $2.7634923 \times 10^{0}$ & $-3.2194242 \times 10^{-5}$ & $1.3333994 \times 10^{0}$ &  &  \\
2 & He & 3 & $1.4779395 \times 10^{-1}$ & $1.3980103 \times 10^{0}$ & $2.0588847 \times 10^{-2}$ & $4.4712824 \times 10^{-1}$ & $2.7061783 \times 10^{-2}$ & $6.8156976 \times 10^{-1}$ \\
3 & Li & 5 & $-1.5391772 \times 10^{0}$ & $6.9424486 \times 10^{0}$ & $1.1463302 \times 10^{-1}$ & $3.0129001 \times 10^{0}$ & $7.3991932 \times 10^{-2}$ & $4.4979492 \times 10^{-1}$ \\
 &  &  & $5.7409070 \times 10^{-3}$ & $1.9060536 \times 10^{-1}$ & $1.9919285 \times 10^{0}$ & $7.2108374 \times 10^{0}$ &  &  \\
4 & Be & 5 & $3.5658514 \times 10^{0}$ & $4.5875611 \times 10^{0}$ & $-1.9486251 \times 10^{0}$ & $3.7887404 \times 10^{0}$ & $4.0966745 \times 10^{-2}$ & $2.3362249 \times 10^{-1}$ \\
 &  &  & $2.7736526 \times 10^{-3}$ & $1.0485891 \times 10^{-1}$ & $-8.9346692 \times 10^{-2}$ & $1.0918026 \times 10^{0}$ &  &  \\
5 & B & 5 & $2.7696788 \times 10^{-1}$ & $5.3045349 \times 10^{0}$ & $1.2250540 \times 10^{0}$ & $3.3803432 \times 10^{0}$ & $2.5167104 \times 10^{-2}$ & $1.3828601 \times 10^{-1}$ \\
 &  &  & $1.1727011 \times 10^{-3}$ & $6.2630609 \times 10^{-2}$ & $-8.7177679 \times 10^{-2}$ & $6.3790417 \times 10^{-1}$ &  &  \\
\bottomrule
\end{longtable}

\endgroup

\ConflictsOfInterest{The authors declare that they have no known competing financial interests or personal relationships that could have appeared to influence the work reported in this paper.}

\DataAvailability{The fitted coefficients for all elements---the Dirac--Pad\'e parameterisation, the parameter-matched non-relativistic baseline, and the fixed five-term refit (`Fixed-5'), in single precision (float32) for direct use in GPU multislice simulation---together with the parameterisation-evaluation implementation, the complete-domain positivity validation script and machine-readable results, and the relativistic reference densities used as the fit targets, are publicly available at \url{https://github.com/NeuralSoftX/scattering_factors_2026}. The complete \texttt{atomx} package is not part of this repository and is maintained separately by Zhang.}

\begingroup
\clubpenalty=10000
\widowpenalty=10000
\bibliography{references}
\endgroup

\end{document}